\documentclass[twocolumn]{aastex62}

\usepackage{amsmath}
\graphicspath{{./}{figures/}}

\received{February 28, 2020}
\revised{April 24, 2020}
\accepted{May 4, 2020}

\submitjournal{ApJ}

\begin{document}

\title{ARES I\footnote{ARES: Ariel Retrieval of Exoplanets School}: WASP-76 b, A Tale of Two HST Spectra}

\correspondingauthor{Billy Edwards}
\email{billy.edwards.16@ucl.ac.uk}

\author{Billy Edwards}
\affil{Department of Physics and Astronomy, University College London, London, United Kingdom}

\author{Quentin Changeat}
\affil{Department of Physics and Astronomy, University College London, London, United Kingdom}

\author{Robin Baeyens}
\affil{Instituut voor Sterrenkunde, KU Leuven, Celestijnenlaan 200D bus 2401, 3001 Leuven, Belgium}

\author{Angelos Tsiaras}
\affil{Department of Physics and Astronomy, University College London, London, United Kingdom}

\author{Ahmed Al-Refaie} 
\affil{Department of Physics and Astronomy, University College London, London, United Kingdom}

\author{Jake Taylor}
\affil{Department of Physics (Atmospheric, Oceanic and Planetary Physics), University of Oxford, Parks Rd, Oxford, United Kingdom}

\author{Kai Hou Yip}
\affil{Department of Physics and Astronomy, University College London, London, United Kingdom}

\author{Michelle Fabienne Bieger}
\affil{College of Engineering, Mathematics and Physical Sciences,  University of Exeter, North Park Road, Exeter, United Kingdom}

\author{Doriann Blain}
\affil{LESIA Observatoir\'e de Paris, Section de Meudon 5, place Jules Janssen 92195 Meudon, France}

\author{Am\'elie Gressier}
\affil{LATMOS, CNRS, Sorbonne Universit\'e UVSQ, 11 boulevard d'Alembert, F-78280 Guyancourt, France}
\affil{Sorbonne Universit\'es, UPMC Universit\'e Paris 6 et CNRS, 
		UMR 7095, Institut d'Astrophysique de Paris, Paris, France}
\affil{LESIA Observatoir\'e de Paris, Section de Meudon 5, place Jules Janssen 92195 Meudon, France}

\author{Gloria Guilluy}
\affil{Dipartimento di Fisica, Universit\'{a} degli Studi di Torino, via Pietro Giuria 1, I-10125 Torino, Italy}
\affil{INAF Osservatorio Astrofisico di Torino, Via Osservatorio 20, I-10025 Pino Torinese, Italy}

\author{Adam Yassin Jaziri}  
\affil{Laboratoire d'astrophysique de Bordeaux, Univ. Bordeaux, CNRS, B18N, all\'{e}e Geoffroy Saint-Hilaire, 33615 Pessac, France}

\author{Flavien Kiefer}
\affil{Sorbonne Universit\'es, UPMC Universit\'e Paris 6 et CNRS, 
UMR 7095, Institut d'Astrophysique de Paris, Paris, France}

\author{Darius Modirrousta-Galian} 
	\affil{INAF Osservatorio Astronomico di Palermo, Piazza del Parlamento 1, I-90134 Palermo, Italy}
	\affil{University of Palermo, Department of Physics and Chemistry, Via Archirafi 36, Palermo, Italy}
	
\author{Mario Morvan}
	\affil{Department of Physics and Astronomy, University College London, London, United Kingdom}

\author{Lorenzo V. Mugnai} 
	\affil{La Sapienza Universit\'a di Roma, Department of Physics, Piazzale Aldo Moro 2, 00185 Roma, Italy}

\author{William Pluriel}  
	\affil{Laboratoire d'astrophysique de Bordeaux, Univ. Bordeaux, CNRS, B18N, all\'{e}e Geoffroy Saint-Hilaire, 33615 Pessac, France}

\author{Mathilde Poveda}
	\affil{Laboratoire Interuniversitaire des Syst\`{e}mes Atmosph\'{e}riques (LISA), UMR CNRS 7583, Universit\'{e} Paris-Est-Cr\'eteil, Universit\'e de Paris, Institut Pierre Simon Laplace, Cr\'{e}teil, France}
	\affil{Maison de la Simulation, CEA, CNRS, Univ. Paris-Sud, UVSQ, Universit\'{e} Paris-Saclay, F-91191 Gif-sur-Yvette, France} 
	
\author{Nour Skaf} 
 	\affil{LESIA Observatoir\'e de Paris, Section de Meudon 5, place Jules Janssen 92195 Meudon, France}
	 \affil{Department of Physics and Astronomy, University College London, London, United Kingdom}

\author{Niall Whiteford}
 \affil{Institute for Astronomy, University of Edinburgh, Blackford Hill, Edinburgh EH9 3HJ, UK}
\affil{Centre for Exoplanet Science, University of Edinburgh, Edinburgh EH9 3FD, UK}

\author{Sam Wright}
\affil{Department of Physics and Astronomy, University College London, London, United Kingdom}

\author{Tiziano Zingales} 
\affil{Laboratoire d'astrophysique de Bordeaux, Univ. Bordeaux, CNRS, B18N, all\'{e}e Geoffroy Saint-Hilaire, 33615 Pessac, France}

\author{Benjamin Charnay}
\affil{LESIA Observatoir\'e de Paris, Section de Meudon 5, place Jules Janssen 92195 Meudon, France}

\author{Pierre Drossart}  
\affil{LESIA Observatoir\'e de Paris, Section de Meudon 5, place Jules Janssen 92195 Meudon, France}

\author{J\'{e}r\'{e}my Leconte}  
\affil{Laboratoire d'astrophysique de Bordeaux, Univ. Bordeaux, CNRS, B18N, all\'{e}e Geoffroy Saint-Hilaire, 33615 Pessac, France}

\author{Olivia Venot}  
\affil{Laboratoire Interuniversitaire des Syst\`{e}mes Atmosph\'{e}riques (LISA), UMR CNRS 7583, Universit\'{e} Paris-Est-Cr\'eteil, Universit\'e de Paris, Institut Pierre Simon Laplace, Cr\'{e}teil, France}

\author{Ingo Waldmann}
\affil{Department of Physics and Astronomy, University College London, London, United Kingdom}

\author{Jean-Philippe Beaulieu}
\affil{School of Physical Sciences, University of Tasmania,
Private Bag 37 Hobart, Tasmania 7001 Australia}
\affil{Sorbonne Universit\'es, UPMC Universit\'e Paris 6 et CNRS, 
UMR 7095, Institut d'Astrophysique de Paris, Paris, France}

\begin{abstract}

We analyse the transmission and emission spectra of the ultra-hot Jupiter WASP-76 b, observed with the G141 grism of the Hubble Space Telescope's Wide Field Camera 3 (WFC3). We reduce and fit the raw data for each observation using the open-source software Iraclis before performing a fully Bayesian retrieval using the publicly available analysis suite TauRex 3. Previous studies of the WFC3 transmission spectra of WASP-76 b found hints of titanium oxide (TiO) and vanadium oxide (VO) or non-grey clouds. Accounting for a fainter stellar companion to WASP-76, we reanalyse this data and show that removing the effects of this background star changes the slope of the spectrum, resulting in these visible absorbers no longer being detected, eliminating the need for a non-grey cloud model to adequately fit the data but maintaining the strong water feature previously seen. However, our analysis of the emission spectrum suggests the presence of TiO and an atmospheric thermal inversion, along with a significant amount of water. Given the brightness of the host star and the size of the atmospheric features, WASP-76 b is an excellent target for further characterisation with HST, or with future facilities, to better understand the nature of its atmosphere, to confirm the presence of TiO and to search for other optical absorbers.

\end{abstract}

\section{Introduction}

Ultra-hot Jupiters are an intriguing population of exoplanets. With dayside temperatures greater than $\sim$2000 K, these planets were truly unexpected and continue to unveil surprising traits. Despite being a rare outcome of planetary formation, many have been found using ground-based surveys such as the Wide Angle Search for Planets (WASP, \citet{pollacco}), the Hungarian Automated Telescope network (HAT, \citet{bakos_south}) and Kilodegree Extremely Little Telescope (KELT, \citet{pepper_kelt}). Given their size and temperature, as well as the brightness of their host stars, these planets are excellent targets for atmospheric characterisation. Moreover, they offer the opportunity to explore atmospheric chemistry and dynamics in extreme conditions. Understanding their composition, and thus their metallicity and carbon to oxygen ratio, is crucial for constraining formation and migration theories \citep{venturini, madhu_formation}.

The Wide Field Camera 3 (WFC3) onboard the Hubble Space Telescope (HST) has, along with Spitzer's InfraRed Array Camera (IRAC), been at the forefront of characterising these planets. These very hot atmospheres were predicted to have inverted temperature-pressure profiles due to strong optical absorption by TiO and VO \citep{Hubeny_thermal_inv, Fortney_2008}. HST observations of two cooler hot Jupiters (T $<$ 2000 K) detected non-inverted temperature profiles for WASP-43\,b \citep{stevenson} and HD\,209458\,b \citep{line_hd209}, which are consistent with the theoretical predictions of \cite{Fortney_2008}. However, observations of the emission spectra of ultra-hot Jupiters have, thus far, been inconclusive on their thermal structure and composition. While some have shown features due to water or optical absorbers, others are consistent with a simple blackbody fit \cite[e.g.][]{madhu_wasp12_spitzer, Haynes_Wasp33b_spectrum_em, beatty_kep13,Evans_wasp121_t1,Evans_wasp121_e1,Evans_wasp121_t2, Arcangeli_2018, Kreidberg_w103, Evans_wasp121_e2,bourrier}. This variety may well be because the emission spectrum in this band-pass is dependent on both the water content and the thermal structure of the planet, and the G141 grism (1.1-1.7 $\mu$m)  probes a region of the atmosphere where the temperature only varies slowly with pressure \citep{Parmentier_2018_w121photodiss, mansfield_hatp7}.
 
Upon the discovery of a companion, WASP-76 became the brightest star known to host a planet with a radius greater than 1.5 R$_J$ \citep{west_w76-82-90}. While brighter targets have since been discovered, WASP-76 b is still one of the best currently-known targets for atmospheric characterisation and the transmission spectrum was observed by the HST in November 2015. The observations were taken with the WFC3 using the G141 grism which covers 1.1-1.7 $\mu m$. This spectrum was analysed by \citet{tsiaras_30planets} as part of a population study of 30 gaseous exoplanets. Retrievals by \citet{tsiaras_30planets} suggested a water rich atmosphere (log(H$_2$O) = $-2.7\pm1.07$) with a 4.4$\sigma$ detection of TiO and VO. However, as noted in the study, the abundances of TiO retrieved were likely to be nonphysical and affected by correlations between the molecular abundances, planet radius and cloud pressure. 

Retrieval analysis of this spectrum was also performed by \cite{fisher}, who extracted a water abundance which was incompatible with the previous study (log(H$_2$O) = $-5.3\pm0.61$). \cite{fisher} did not fit for TiO or VO but instead used a non-grey cloud model to explain the opacity seen at shorter wavelengths. WASP-76 b was one of two planets from their study of 38 transmission spectra that was not well-fitted using the standard grey cloud model. 

High resolution ground-based observations offer the opportunity to resolve the spectral lines of exoplanet atmospheres and the High Accuracy Radial velocity Planet Searcher (HARPS, \citet{pepe_harps}) was used to analyse WASP-76b and find evidence for absorption due to sodium \citep{seidel,zak}. Recent work by \cite{von_essen_w76} using transmission data from Hubble's Space Telescope Imaging Spectrograph (STIS) confirmed the presence of sodium and provided marginal evidence of titanium hydride (TiH).

Two transits of WASP-76 b were observed using the Echelle Spectrograph for Rocky Exoplanets and Stable Spectroscopic Observations (ESPRESSO) instrument on the Very Large Telescope (VLT) and used to reveal an asymmetric absorption signature which was attributed to neutral iron (Fe, \cite{ehrenreich_wasp76}). The signature was blue-shifted on the trailing limb, proving evidence for strong day-to-night winds and an asymmetric dayside. However, the lack of signal on the leading limb means little Fe is present there and thus must be condensing on the nightside.

Here we present an analysis of the transmission and emission spectra of WASP-76\,b, taken with the WFC3 G141 grism aboard HST. Although not reported in the discovery paper \citep{west_w76-82-90}, the presence of a stellar companion was noted in several studies \citep{wollert, ginski, ngo_w76, bohn_w76} and the common proper motion of the two objects was confirmed by \cite{southworth_w76}. This was not accounted for in previous WFC3 transmission studies and, using Wayne simulations \citep{varley}, we show that this companion affects the spectral data obtained. We use Wayne to remove the contamination of the companion and reanalyse the transmission spectrum finding evidence for H$_2$O but, while we were able to place upper limits on the TiO, VO, and FeH abundances, we were unable to well constrain other molecules. In the emission spectrum, we find indications of the presence of H$_2$O and TiO, along with a thermal inversion. Additionally we place upper limits on the abundances of FeH and VO.

\section{Data Analysis}
\subsection{Data Reduction}

Our analysis started from the raw spatially scanned spectroscopic images which were obtained from the Mikulski Archive for Space Telescopes\footnote{\url{https://archive.stsci.edu/hst/}}. The transmission spectrum was acquired as part of proposal 14260, taken in November 2015, while the observation of the eclipse was taken during proposal 14767 in November 2016. We used Iraclis\footnote{\url{https://github.com/ucl-exoplanets/Iraclis}}, a specialised, open-source software for the analysis of WFC3 scanning observations \citep{tsiaras_hd209} and the reduction process included the following steps: zero-read subtraction, reference pixels correction, non-linearity correction, dark current subtraction, gain conversion, sky background subtraction, calibration, flat-field correction, and corrections for bad pixels and cosmic rays. For a detailed description of these steps, we refer the reader to the original Iraclis paper \citep{tsiaras_hd209}. 

\begin{table}
    \centering
    \begin{tabular}{cc}\hline\hline
    \multicolumn{2}{c}{Input Stellar \& Planetary Parameters$^*$}\\ \hline\hline
    $T_{*}$ [K] & 6329$\pm$65 \\
    $R_{*}$ [$R_\odot$] & 1.756$\pm$0.071 \\
    $M_{*}$ [M$_\odot$]  & 1.458$\pm$0.021 \\
    $\log_{10}(g)$ [$cm/s^{2}$] & 4.196$\pm$0.106\\
    $[\text{Fe/H}]$ &  0.366$\pm$0.053\\
    a/R$_*$ & 4.08$^{+0.02}_{-0.06}$\\
    $e$ & 0 (fixed)\\
    $i$ &  89.623 $^{+0.005}_{-0.034}$\\
    $\omega$ &  0 (fixed) \\
    $P$ [days] & 1.80988198$^{+0.00000064}_{-0.00000056}$\\
    $T_0$ [$BJD_{TDB}$] & 2458080.626165$^{+0.000418}_{-0.000367}$\\
    $M_{p}$ &  0.894$^{+0.014}_{-0.013}$\\
    $R_{p}$ & 1.854$^{0.077}_{-0.076}$ \\ \hline
        \multicolumn{2}{c}{$^*$Taken from \cite{ehrenreich_wasp76}}\\\hline \hline
    \multicolumn{2}{c}{Companion Star Parameters}\\\hline \hline
    $T_{*}$ [K] & 4824$^\dagger$\\
    $R_{*}$ [$R_\odot$] & 0.83$^\dagger$\\
    $M_{*}$ [M$_\odot$]  & 0.79$^\dagger$\\
    $\log_{10}(g)$ [$cm/s^{2}$]  & 4.5$^\ddagger$\\
    $[\text{Fe/H}]$ &  0.0$^\ddagger$\\
    
    \hline \hline

    \multicolumn{2}{c}{$^\dagger$Taken or derived from \cite{bohn_w76}}\\
    \multicolumn{2}{c}{$^\ddagger$Assumed value}\\\hline\hline
    \end{tabular}
    \caption{Stellar and planetary parameters for WASP-76\,b used during the Iraclis, Wayne and TauREx analyses.}
    \label{tab:star_params}
\end{table}

\begin{figure}
    \centering
    \includegraphics[width=0.475\textwidth]{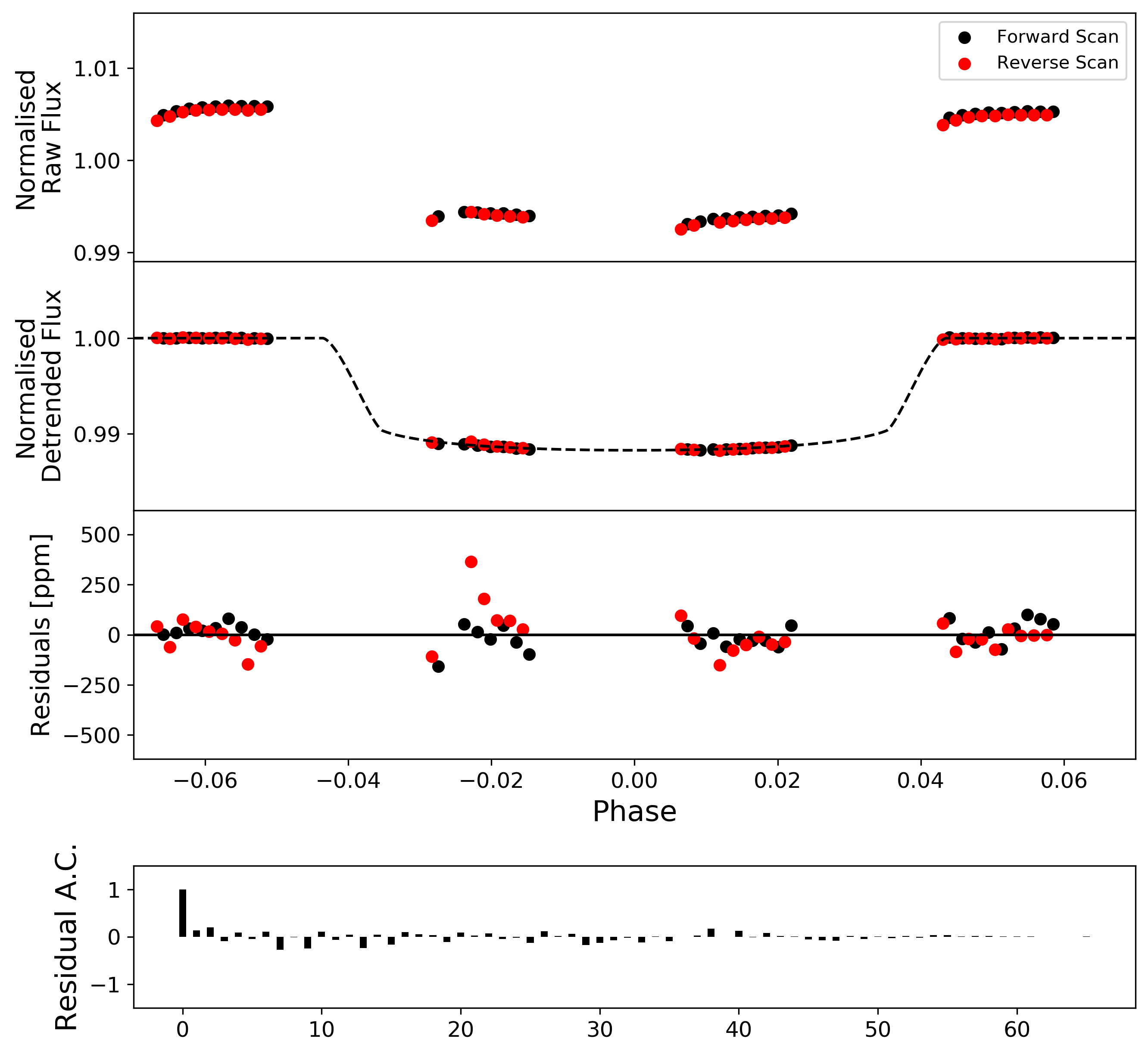}
    \includegraphics[width=0.475\textwidth]{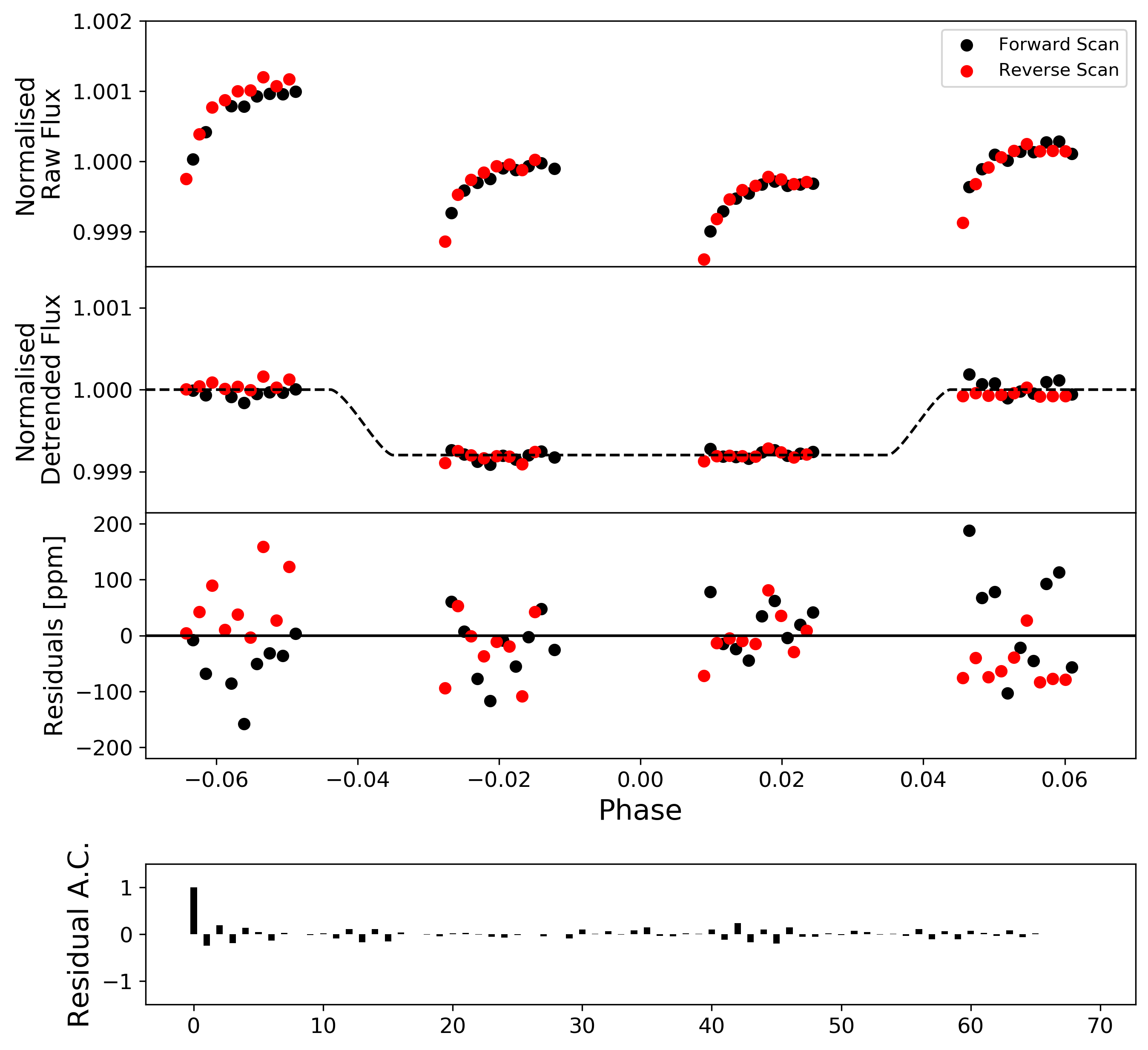}
    \caption{White light curves for the transmission (top) and emission (bottom) observations of WASP-76 b. First panel: raw light curve, after normalisation. Second panel: light curve, divided by the best fit model for the systematics. Third panel: residuals for best-fit model. Fourth panel: auto-correlation function of the residuals.}
    \label{fig:white}
\end{figure}

\begin{figure*}
    \centering
    \includegraphics[width=0.8\textwidth]{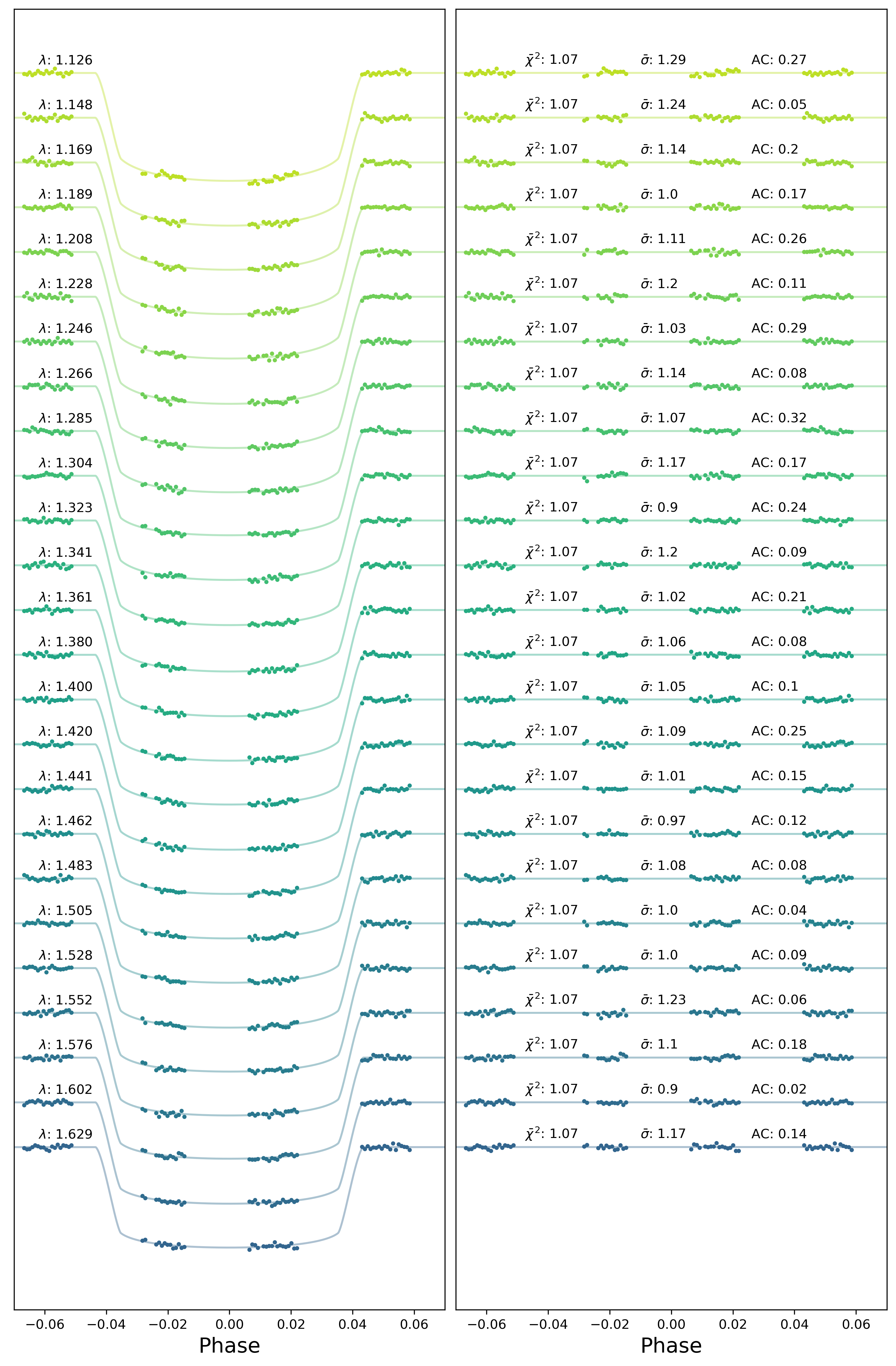}
    \caption{Spectral light curves fitted with Iraclis for the transmission spectra where, for clarity, an offset has been applied. Left: the detrended spectral light curves with best-fit model plotted. Right: residuals from the fitting with values for the Chi-squared ($\chi^2$), the standard deviation of the residuals with respect to the photon noise ($\bar{\sigma}$) and the auto-correlation (AC).}
    \label{fig:transit_spectral}
\end{figure*}

\begin{figure*}
    \centering
    \includegraphics[width=0.8\textwidth]{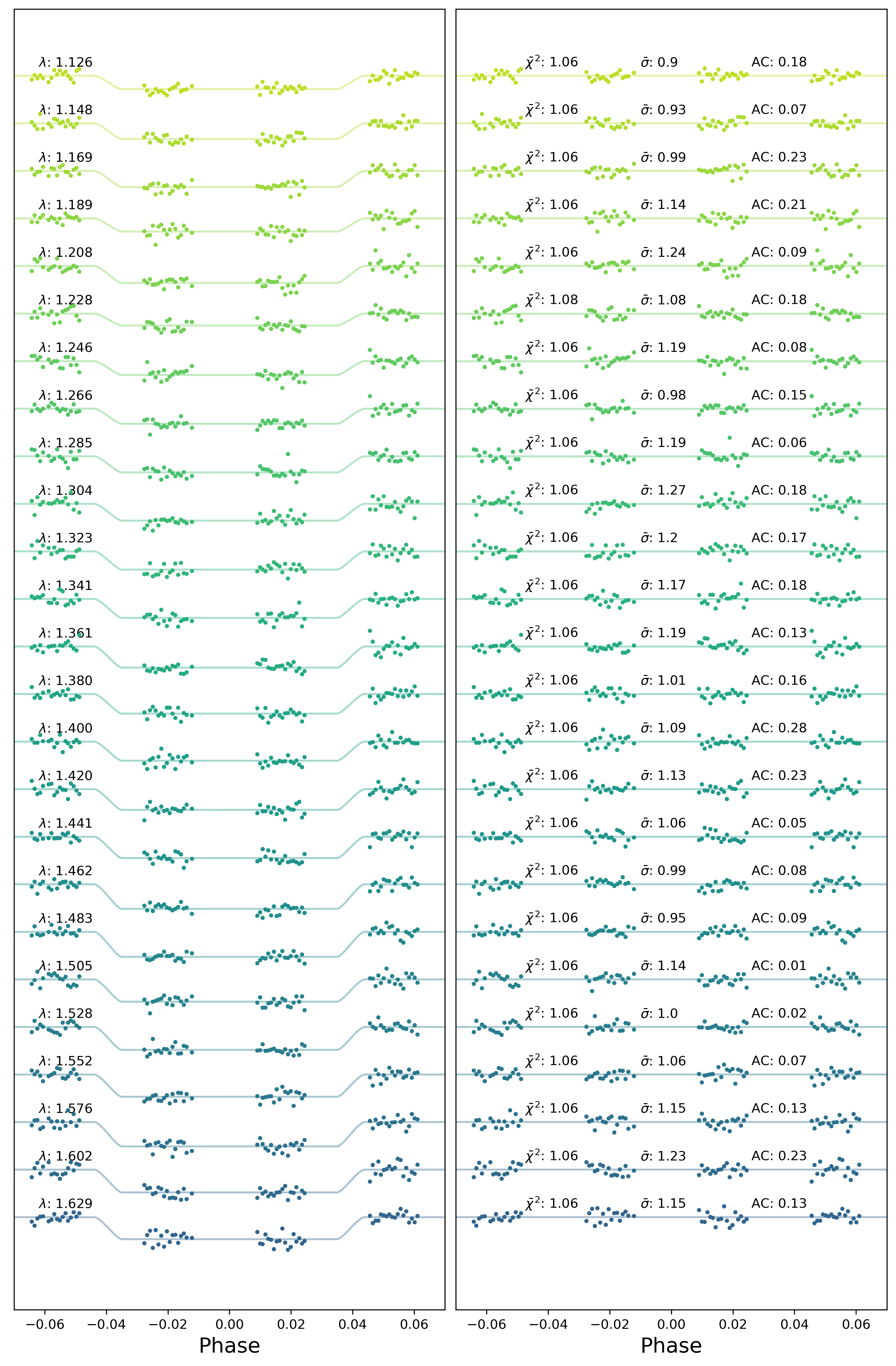}
    \caption{Spectral light curves fitted with Iraclis for the emission spectra where, for clarity, an offset has been applied. Left: the detrended spectral light curves with best-fit model plotted. Right: residuals from the fitting with values for the Chi-squared ($\chi^2$), the standard deviation of the residuals with respect to the photon noise ($\bar{\sigma}$) and the auto-correlation (AC).}
    \label{fig:eclipse_spectral}
\end{figure*}

The reduced spatially scanned spectroscopic images were then used to extract the white (from 1.1-1.7 $\mu$m) and spectral light curves. The spectral light curves bands were selected such that the SNR is approximately uniform across the planetary spectrum. We then discarded the first orbit of each visit as they present stronger wavelength-dependent ramps, and the first exposure after each buffer dump as these contain significantly lower counts than subsequent exposures \citep[e.g.][]{deming_hd209, tsiaras_hd209, mansfield_hatp7}. Additionally, for the third orbit of the transit, two further scans were removed to increase the quality of the fit \citep[e.g.][]{Kreidberg_w103}. For the fitting of the white light curves, the only free parameters were the mid-transit time and planet-to-star ratio. \citet{alexoudi_inc} showed that the inclination can have a strong effect on the derived slope of optical transmission data. We therefore checked that our results were not affected in this way by running light curve fittings with the inclination as a free parameter. The spectral transit depths from these fittings did not differ from the fixed inclination case. However, the best-fit inclinations differed between the transit and eclipse light curve and so the orbital parameters were set to values from \citet{ehrenreich_wasp76}. The limb-darkening coefficients were selected from the best available stellar parameters using values from \citet{claretI,claretII} and using the stellar parameters from \citet{ehrenreich_wasp76}. We did not fit for the limb darkening coefficients as they are degenerate with other parameters, particularly given the periodic gaps in the HST data. \citet{tsiaras_30planets} showed that fitting the  limb darkening coefficients doesn't generally affect the recovered spectrum and \citet{morello_ldc} showed that uncertainties in stellar models do not significantly affect the atmospheric spectra in the WFC3 spectral band. The fitted white light curves for both observations are shown in Figure \ref{fig:white} while the spectral light curves are plotted in Figures \ref{fig:transit_spectral} and  \ref{fig:eclipse_spectral}.

\subsection{Removal of Companion Contamination}

The stellar companion of WASP-76, reported by \citet{bohn_w76} and \citet{ southworth_w76} has a K magnitude which is $\sim$2.30 fainter and the separation between both stars is only 0.436''. As such it is expected to have contaminated the transmission and emission spectra obtained by Hubble. For exoplanet spectroscopy, this third light modifies the transit/eclipse depth. For the Hubble STIS observations analysed by \cite{von_essen_w76}, the point spread functions (PSFs) of WASP-76 and its companion could be distinguished. However, due to the plate scale of HST WFC3 it is not resolvable in this case. Hence, to account for this, we used the freely available WFC3 simulator Wayne\footnote{\url{https://github.com/ucl-exoplanets/wayne}}.

\begin{figure}
    \centering
    \includegraphics[width=0.475\textwidth]{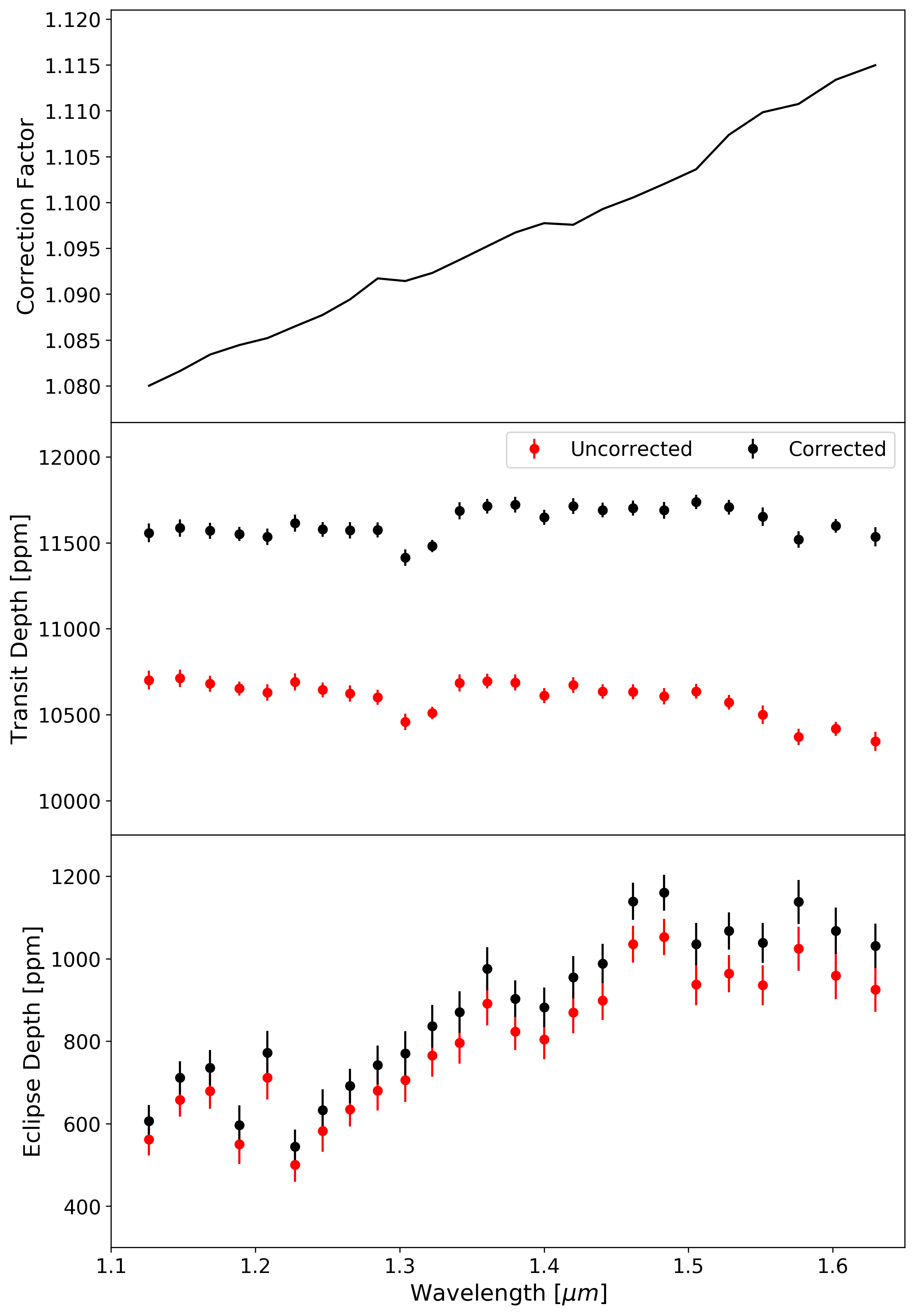}
    \caption{Top: Wavelength-dependent correction factor derived from the Wayne simulations. Middle: the corrected (black) and uncorrected (red) transmission spectra. Bottom: the same but for the emission spectra.}
    \label{fig:correction}
\end{figure}

Wayne is capable of producing grism spectroscopic frames, both in staring and in spatial scanning modes \citep{varley}. Using the stellar parameters from \cite{bohn_w76}, we utilised Wayne to model the contribution of the companion star to the spectral data obtained. We created simulated detector images of both the main and companion star, using these to extract the flux contribution in each spectral bin of each star. The correction to the spectra is then applied as a wavelength dependant dilution factor which is derived as a ratio of extracted flux between the stars. Such an approach has previously been used on WFC3 data (e.g. for WASP-12 b \cite{stevenson_wasp12, Kreidberg_wasp12, tsiaras_30planets}). The recovered transmission and emission spectra, before and after the correction was applied, are shown in Figure \ref{fig:correction} along with the correction factor used. The values are also given in Table \ref{tab:spectra}, along with the corrected transmission and emission depths. Two trends are seen: firstly, the transit and eclipse depths are increased and, secondly, the slope of the spectrum is changed in each case due to the differing spectral types of the stars.

\subsection{Atmospheric Modelling}

The retrieval of the transmission and emission spectrum were performed using the publicly available retrieval suite TauREx 3 \citep{al-refaie_taurex3}\footnote{\url{https://github.com/ucl-exoplanets/TauREx3_public}}. For the star parameters and the planet mass, we used the values from \cite{ehrenreich_wasp76} listed in Table \ref{tab:star_params}. In our runs we assumed that WASP-76 b possesses a primary atmosphere with a ratio H$_2$/He = 0.17. To this we added trace gases and included the molecular opacities from the ExoMol \citep{Tennyson_exomol}, HITRAN \citep{gordon} and HITEMP \citep{rothman} databases for: H$_2$O \citep{polyansky_h2o}, CH$_4$ \citep{exomol_ch4}, CO \citep{li_co_2015}, CO$_2$ \citep{rothman_hitremp_2010}, FeH \citep{dulick_FeH, wende_FeH}, TiO \citep{McKemmish_TiO_new}, VO \citep{mckemmish_vo} and H-. On top of this, we also included Collision Induced Absorption (CIA) from H$_2$-H$_2$ \citep{abel_h2-h2, fletcher_h2-h2} and H$_2$-He \citep{abel_h2-he} as well as Rayleigh scattering for all molecules. 

For the H- opacity, we used the description in \citet{john_1988_h-}. The bound-free absorption coefficient ($k^{bf}(\lambda, T)$) corresponds to the photo-detachment of an electron by hydrogen ion and the free-free absorption coefficient ($k^{ff}(\lambda, T)$) results from the interaction of free electrons in the field of neutral hydrogen atoms. These coefficients (in cm$^4$ dyne$^{-1}$) are expressed per unit electron pressure and per hydrogen atom. One can calculate the electron partial pressure in dyne cm$^{-2}$ ($P_{e-}[\textrm{dyne cm}^{-2}]$) using:
\begin{equation}
    P_{e-}[\textrm{dyne cm}^{-2}] = P [\textrm{bar}]\times V_{e-} \times 10^{6},
\end{equation}
where $P [\textrm{bar}]$ is the atmospheric pressure in bar and $V_{e-}$ is the volume mixing ratio of electrons.
The weighted cross section $\sigma_{H-}$ for the H- absorption is given by:
\begin{equation}
\begin{aligned}
    \sigma_{H-}(\lambda, T) = {} & (k^{bf}(\lambda, T) + k^{ff}(\lambda, T)) \\
                                & \times P_{e-}[\textrm{dyne cm}^{-2}] * V_{H},
\end{aligned}
\end{equation}
where $V_{H}$ is the volume mixing ratio of neutral hydrogen atoms.
Hence we are left with two free parameters: the electron and neutral hydrogen volume mixing ratios. In our retrieval analysis, we fixed the hydrogen volume mixing ratio and imposed a profile inspired from \cite{Parmentier_2018_w121photodiss}. We used the two-layer model from \cite{changeat} to describe the increasing abundance of neutral hydrogen atoms with altitude. We chose a surface abundance of $10^{-2}$, a top abundance of $0.5$ and a layer pressure change at $10^{-1}$ bar. Therefore, the only remaining parameter to constrain the H- absorption is the electron volume mixing ratio $V_{e-}$.

Since in emission spectroscopy the radius is degenerate with temperature \citep[e.g.][]{griffin_degen}, we fixed its value to the best fit value from the transmission retrieval. In transmission we assumed an isothermal atmosphere while for the emission we used a non-physically informed approach consisting of 3 temperature points. This led to 5 free variables: surface temperature ($T_{surf}$), temperature of point 1 ($T_1$), temperature of point 2 ($T_{top}$), pressure of point 1 ($P_1$) and pressure of point at the top ($P_{top}$). These points were allowed to vary freely in the pressure grid ranging from $10$ bar to $10^{-10}$ bar. In our retrieval analysis, we used uniform priors for all parameters as described in Table \ref{tab:retrieval_params}. Finally, we explored the parameter space using the nested sampling algorithm Multinest \citep{Feroz_multinest} with 750 live points and an evidence tolerance of 0.5.

\section{Results} \label{sec:results}

\begin{figure*}
    \centering
    \includegraphics[width=0.975\textwidth]{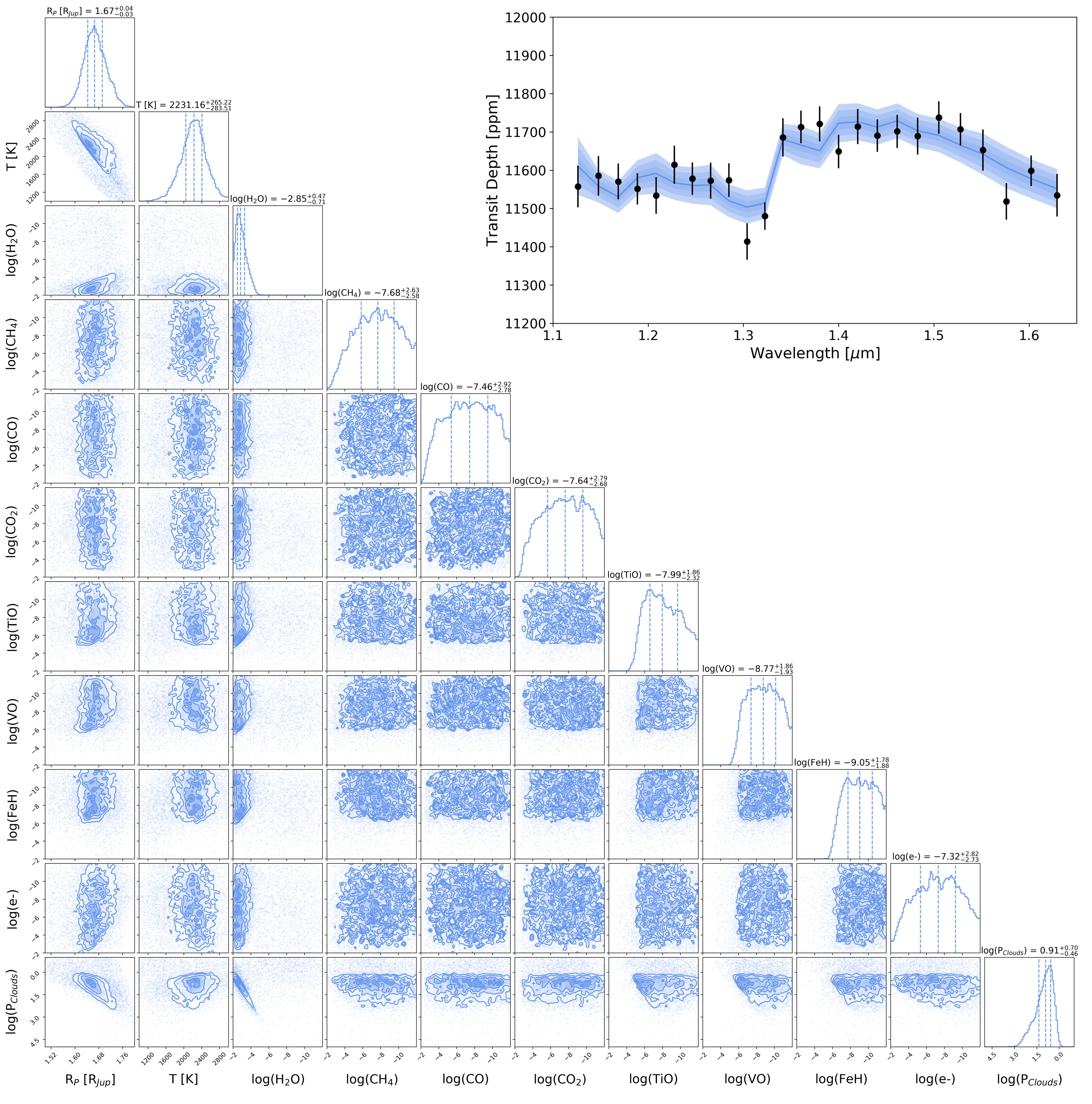}
    \caption{Posterior distributions for the transmission spectrum of WASP-76 b which suggest the presence of a large amount of H$_2$O as well as placing upper limits on the abundances of TiO, VO and FeH. Inset: transmission spectrum (black) with best-fit model and 1-3$\sigma$ uncertainties (blue).}
    \label{fig:wasp79_trans_post}
\end{figure*}

The analysis of the transmission spectra by \citet{tsiaras_30planets} detected water along with the suggestion of TiO and VO. Having accounted for the stellar companion, the slope at the blue end of the spectrum is reduced and thus our retrieval does not find substantial evidence of significant abundances of TiO or VO. However, the recovered water abundance of log(H$_2$O) = -2.85$^{+0.47}_{-0.71}$ is consistent with that from \citet{tsiaras_30planets} (log(H$_2$O) = -2.70$\pm$1.07). While we did attempt to retrieve the carbon-based molecules, CO, CO$_2$, and CH$_4$, we were unable to constrain their abundance as they lack strong features in the G141 wavelength range. Additionally, the abundance of e- (H- opacity) was not constrained but we could place a 1$\sigma$ upper limit of log(FeH)$<$-7.3. Our best-fit model favours the presence of clouds at log(P) = 0.91 Pa but we note there is significant correlation with the abundance of water. The best-fit spectrum and the posteriors are given in Figure \ref{fig:wasp79_trans_post} while the priors and results from the retrieval are given in Table \ref{tab:retrieval_params}. To understand the statistical significance of our results, we also ran a ``molecule free" retrieval where the only fitted parameters were the planet radius, planet temperature and cloud-top pressure. Scattering due to Rayleigh and CIA were also included. The difference in Bayesian log evidence was computed to be 24.7 in favour of the fit including molecules, providing significant evidence of the detection of molecular features ($>$7$\sigma$, \citet{Kass1995bayes}). This is equivalent to the Atmospheric Detectability Index (ADI) as defined in \citet{tsiaras_30planets}.

\begin{figure*}
    \centering
    \includegraphics[width=0.975\textwidth]{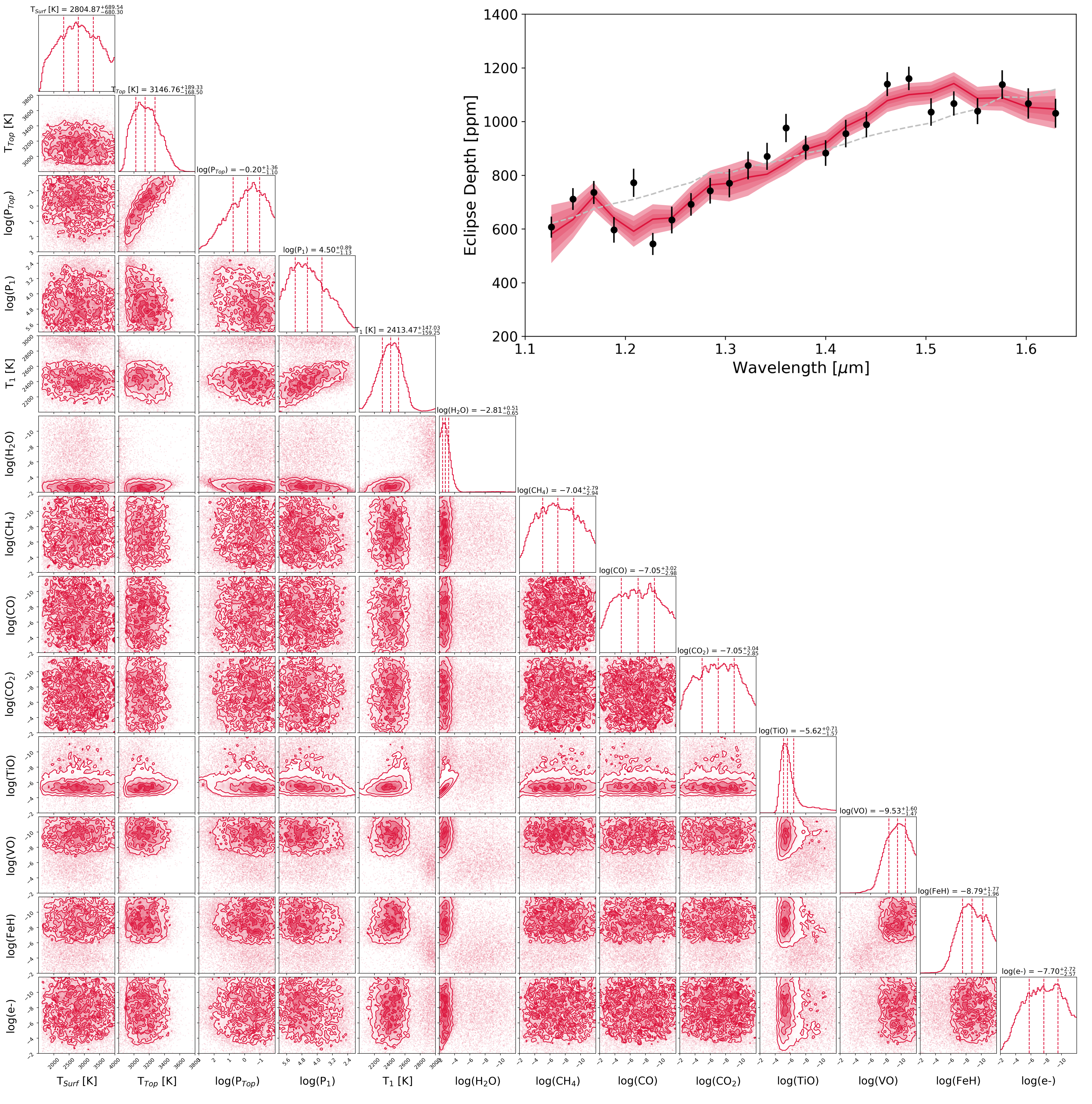}
    \caption{Posterior distributions for the emission spectrum of WASP-76 b which suggest the presence of a large amount of H$_2$O as well as TiO. Inset: emission spectrum (black) with best-fit model and 1-3$\sigma$ uncertainties (red). Also shown is a blackbody fit (grey) which has a temperature of T$_{BB}$ = 2778$\pm$8 K.}
    \label{fig:wasp79_emiss_post}
\end{figure*}

Our retrieval analysis of the emission spectrum of WASP-76 b finds significant evidence of TiO, with an abundance of log(TiO) = -5.62$^{+0.71}_{-1.57}$, along with H$_2$O at a concentration of log(H$_2$O) = -2.81$^{+0.51}_{-0.65}$.  Additionally the emission spectrum places an upper bound on the presence of both iron hydride and vanadium oxide at log(FeH), log(VO) $\approx$ -7. Again the carbon-based molecules were not constrained since there is a lack of spectral information in the WFC3 G141 wavelength band. Due to the presence of optical absorbers, our analysis suggests a temperature inversion in the dayside of WASP-76 b. The retrieval posteriors for the emission spectrum are shown in Figure \ref{fig:wasp79_emiss_post} while Figure \ref{fig_eq_chem} displays the best-fit temperature profiles for both observations. Here the model was compared to a simple blackbody fit, which converged to T$_{BB}$ = 2778$\pm$8 K. The difference in Bayesian log evidence was 12.4, signifying the fit with H$_2$O, TiO and a thermal inversion is statistically preferable at $>$5$\,\sigma$.

\begin{table}
\label{tab:retrieval_params}
\centering
\begin{tabular}{cccc}
\hline\hline
\multicolumn{4}{c}{Transmission}\\ \hline
Parameters & Prior bounds & Scale & Retrieved\\
\hline 
H$_2$O & -12 ; -2 & log  & -2.85 $^{+0.42}_{-0.71}$\\
CH$_4$ & -12 ; -2 & log  &  unconstrained\\
CO & -12 ; -2 & log & unconstrained\\
CO$_2$ & -12 ; -2 & log  & unconstrained\\
TiO & -12 ; -2 & log  & $<$-6.1\\
VO & -12 ; -2 & log  & $<$-6.9\\
FeH & -12 ; -2 & log  & $<$-7.3\\
e- & -12 ; -2 & log  & unconstrained\\
$T_{term}$ (K)& 1600 ; 4000 & linear  & 2231$^{+265}_{-283}$\\
$P_{clouds}$ (Pa)& 6 ; -2 & log  & 0.91$^{+0.70}_{-0.46}$ \\
$R_p$ (R$_{jup}$)& 1.3 ; 2.2 & linear  &  1.67$^{+0.04}_{-0.03}$\\ \hline\hline

\multicolumn{4}{c}{Emission} \\ \hline
Parameters & Prior bounds & Scale & Retrieved\\\hline 
H$_2$O & -12 ; -2 & log & -2.81 $^{+0.51}_{-0.65}$ \\
CH$_4$ & -12 ; -2 & log & unconstrained \\
CO & -12 ; -2 & log  & unconstrained \\
CO$_2$ & -12 ; -2 & log & unconstrained \\
TiO & -12 ; -2 & log  & -5.62 $^{+0.71}_{-1.57}$\\
VO & -12 ; -2 & log  & $<$-7.9\\
FeH & -12 ; -2 & log & $<$-7.0\\
e- & -12 ; -2 & log  & unconstrained \\
$T_{surf}$ (K)& 1600 ; 4000 & linear  & 2805 $^{+689}_{-680}$\\
$T_{1}$ (K)& 1600 ; 4000 & linear   & 2413 $^{+147}_{-159}$\\
$T_{top}$ (K)& 1600 ; 4000 & linear  & 3147 $^{+189}_{-168}$\\
$P_{1}$ (Pa) & 6 ; 2 & log   & 4.50 $^{+0.89}_{-1.13}$\\
$P_{top}$ (Pa) & 3 ; -2 & log  & -0.20 $^{+1.36}_{-1.10}$\\

\hline\hline  \\
\end{tabular}
\caption{List of the retrieved parameters, their uniform prior bounds, the scaling used and the retrieved value.}
\end{table}

\section{Discussion} \label{sec:dis}
\subsection{Comparison to Chemical Models}
To provide context to our findings, we compare the results of our retrieval analysis to a self-consistent forward model computed with petitCODE, a 1D numerical iterator solving for radiative-convective and chemical equilibrium \citep{molliere_petitcode, molliere_jwst}. The code includes radiative scattering, opacities for H$_2$, H$^-$, H$_2$O, CO, CO$_2$, CH$_4$, HCN, H$_2$S, NH$_3$, OH, C$_2$H$_2$, PH$_3$, SiO, FeH, Na, K, Fe, Fe$^+$, Mg, Mg$^+$, TiO and VO, as well as collision induced absorption by H$_2$--H$_2$ and H$_2$--He. Cloud condensation of refractory species is included in the equilibrium chemistry, but no cloud opacities are considered in this simulation. Our petitCODE model for WASP-76 b was computed using the stellar and planetary parameters determined by \cite{west_w76-82-90}. An intrinsic temperature of 600K was adopted for this inflated planet, following the prescription by \cite{thorngren_2019}. A global planetary averaged redistribution of the irradiation was assumed. 

\begin{figure*}
    \centering
    \includegraphics[width= \columnwidth]{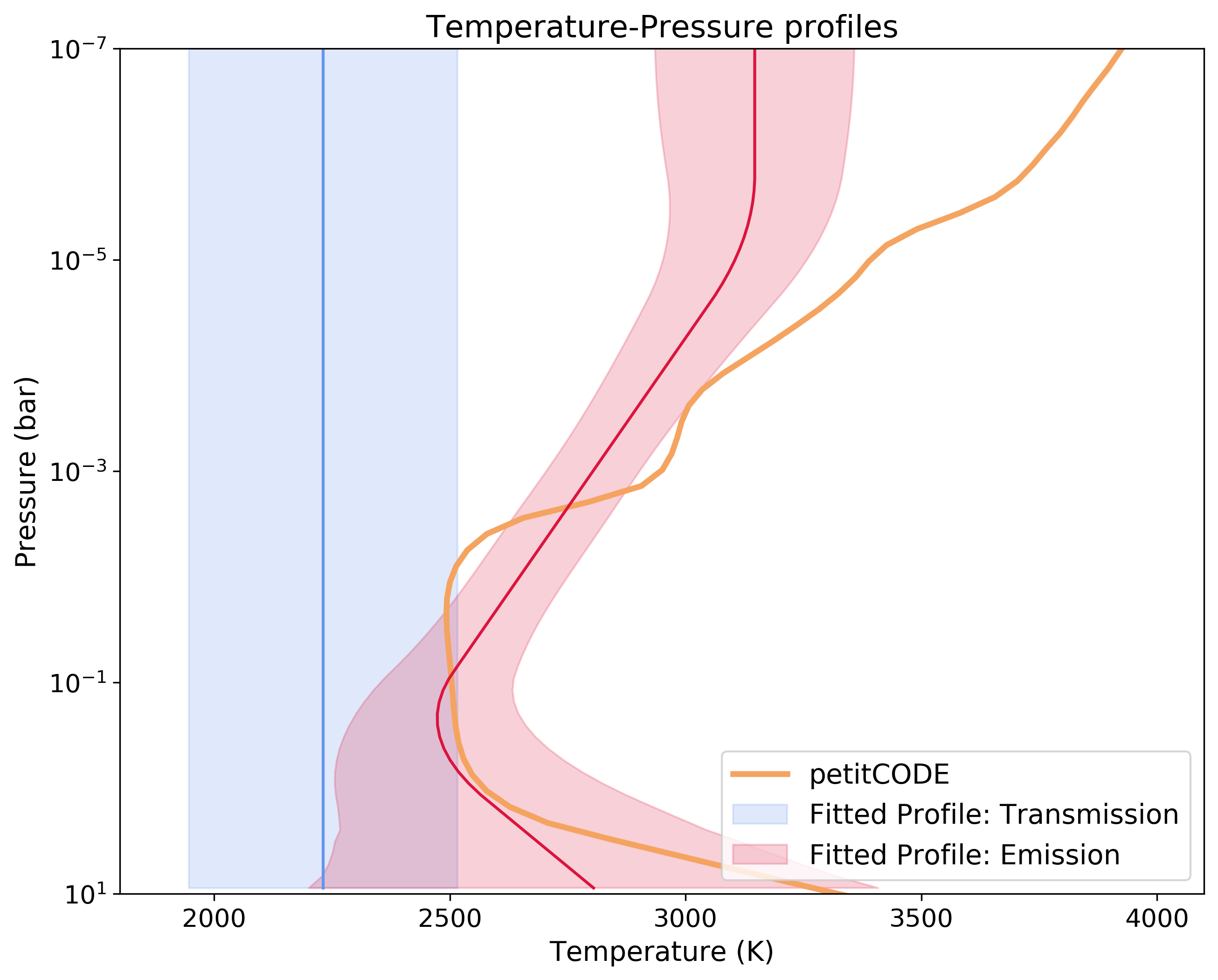}
    \includegraphics[width = \columnwidth]{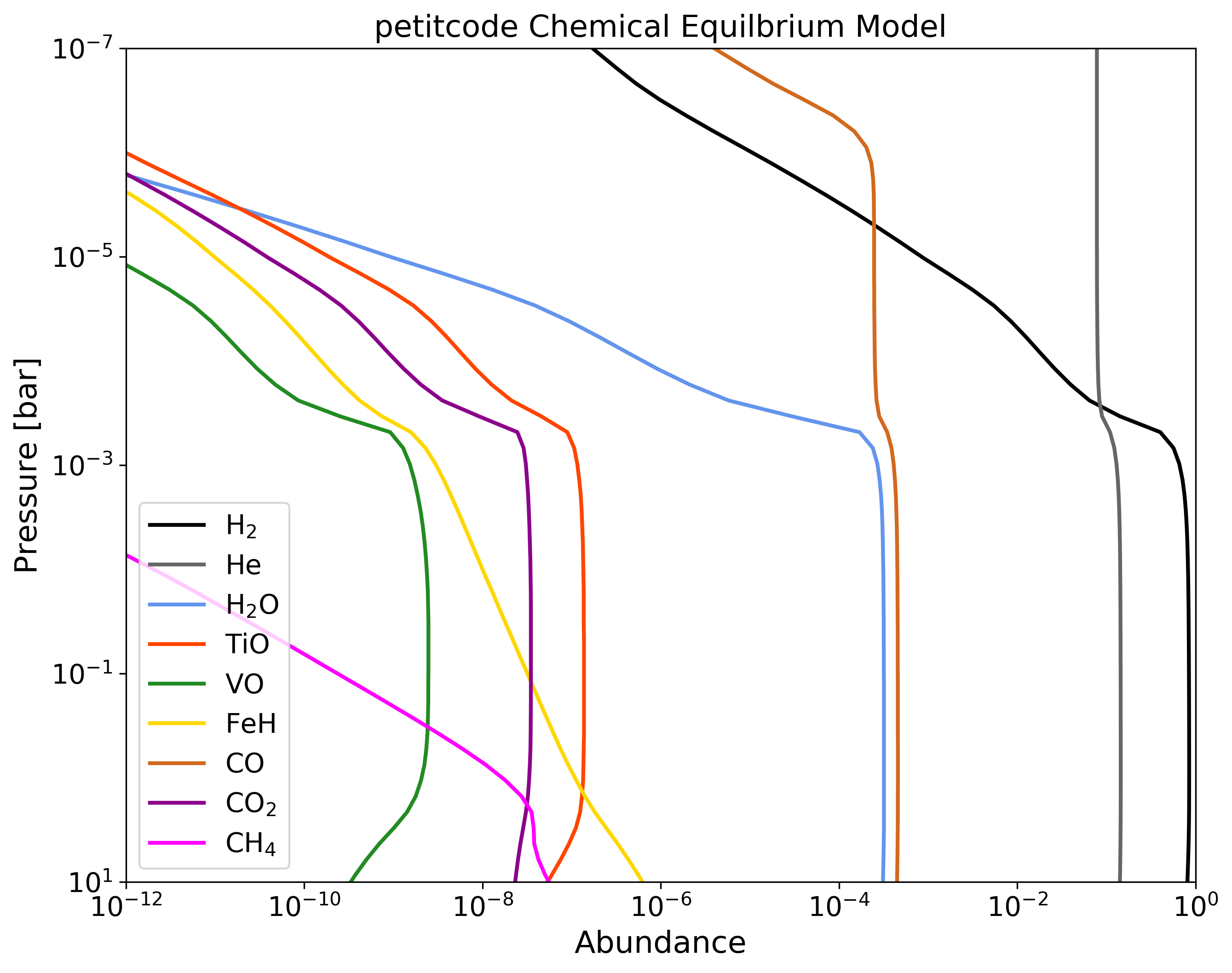}
    \includegraphics[width = \columnwidth]{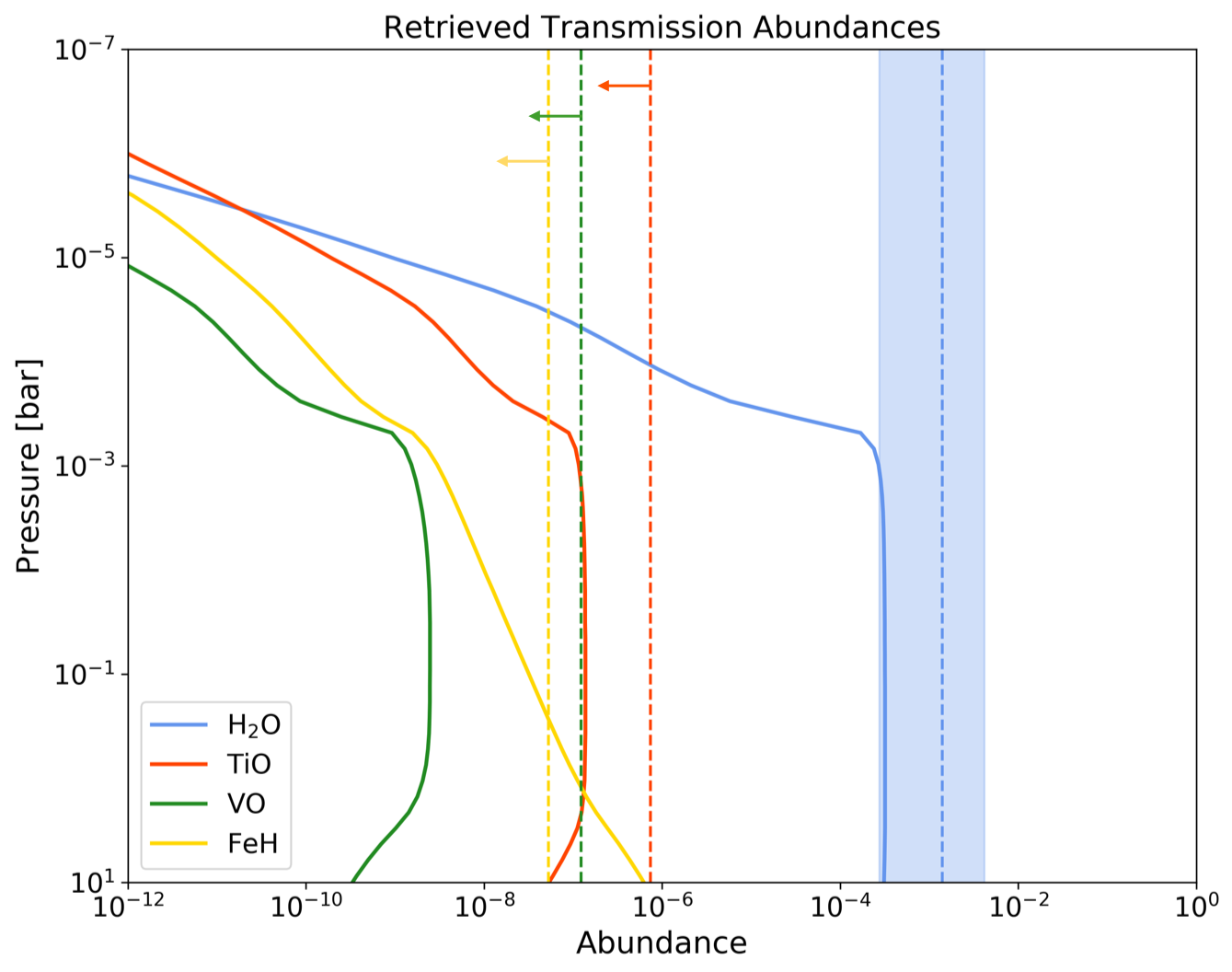}
    \includegraphics[width = \columnwidth]{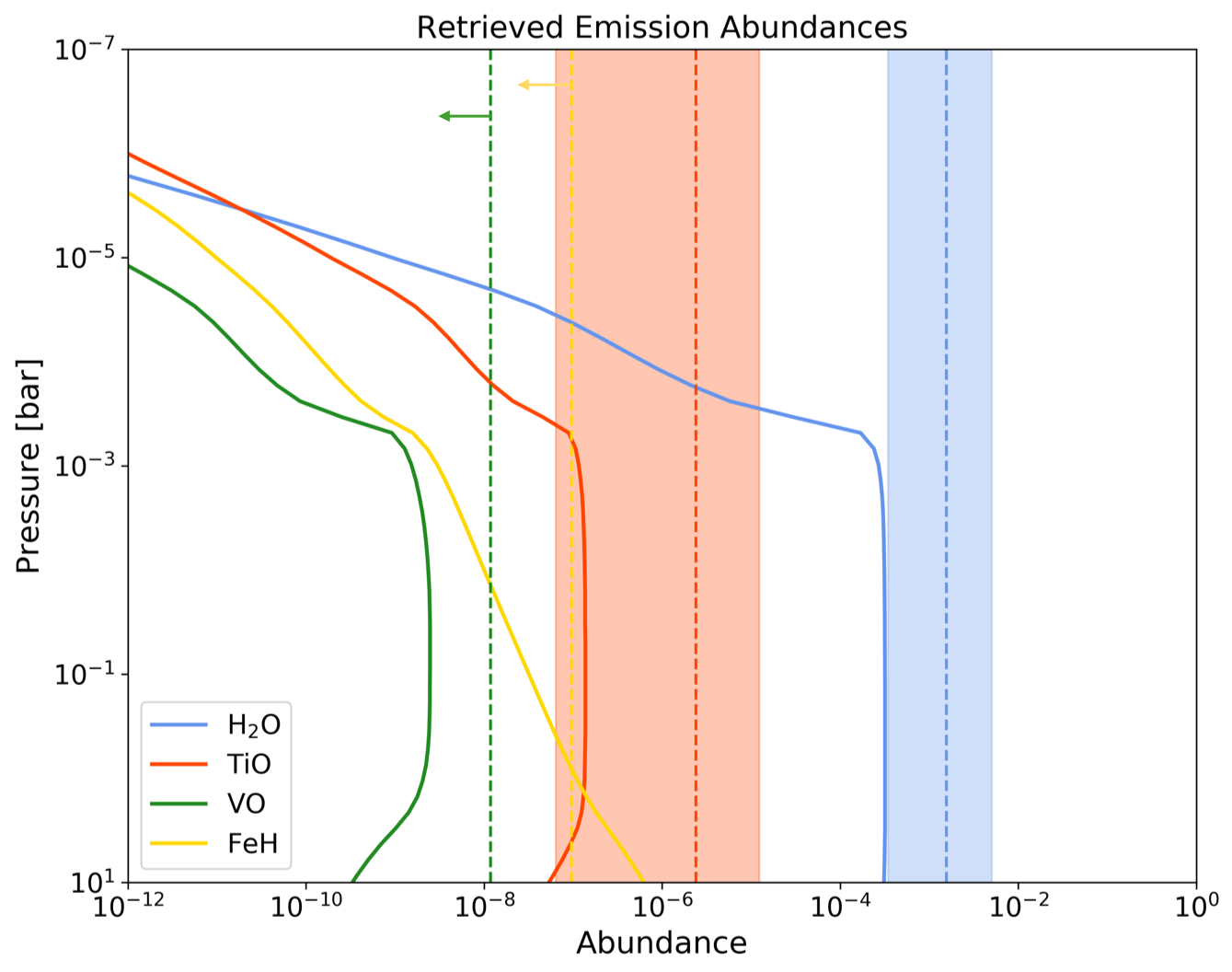}
    \caption{Results of our self-consistent petitCODE model for WASP-76 b and our retrievals on WFC3 data. Top left: comparison of the temperature-pressure profiles. The petitCODE model
    (orange) features a thermal inversion at 1 mbar, due to absorption by TiO and VO, and closely matches the retrieved profile. Top right: Molecular abundances for the petitCODE simulation. The equilibrium fractions of most molecules \textit{(bottom)} remain approximately constant for pressures higher than a few mbar. They drop quickly at lower atmospheric pressures due to thermal dissociation. Bottom left: Comparison of constrained molecular abundances in transmission (dotted lines) to those from the petitCODE simulation (solid lines). The water abundance is seen to be around 1$\sigma$ higher than predicted. Bottom right: Comparison of constrained molecular abundances in emission to those from the petitCODE simulation. Again the water abundance is seen to be around 1$\sigma$ higher than predicted while the TiO concentration is within 1$\sigma$ of the model.}
    \label{fig_eq_chem}
\end{figure*}{}

Two models were produced: one with, and one without, the presence of TiO and VO. For the former case, the resulting temperature-pressure profile and equilibrium abundances are shown in Figure \ref{fig_eq_chem}. The temperature profile shows a inversion in the range probed by transmission/emission spectroscopy (typically $\sim 1$~mbar to $\sim100$~mbar) and this was only present for the model with TiO/VO opacities. A similar result is shown in \citet{lothringer}, who found the same dichotomy between atmospheres with and without TiO/VO for planets with equilibrium temperatures of $T_{eq}$ = 2250 K.

Our retrieval emission abundance for TiO (log(TiO) = -5.62$^{+0.71}_{-1.57}$ is consistent to 1$\sigma$ with the log(TiO) $\approx$ -7 predicted by petitCODE and the upper boundary of log(FeH) $<$ 7 that was retrieved from the emission spectrum also agrees well with the self-consistent petitCODE model. In transmission, the upper limit placed on these molecules is greater than the predicted abundances and thus the non-detection is likely due to the quality of the data. Furthermore, the extent of H$_2$O in the terminator and on the dayside is also similar to the abundance predicted with the chemical equilibrium model. Finally, the VO chemical profile is seen to be below the sensitivity of the emission spectrum. This is due to TiO, FeH and VO possessing similar features in the G141 waveband. This degeneracy may also affect the TiO abundance retrieved.

The equilibrium chemical abundances of most molecules, except CO, drop significantly for pressures lower than a few mbar due to thermal dissociation in the upper atmosphere \citep{lothringer, Parmentier_2018_w121photodiss, Arcangeli_2018}. Models by \cite{Parmentier_2018_w121photodiss} suggest that, for WASP-76 b, nearly half the water should be dissociated at the 1.4 $\mu m$ photosphere. Thus the H$_2$O bands are significantly muted due to thermal dissociation in the upper layers of the atmosphere, owing to the intense irradiation by the nearby host star, as seen in \cite{Arcangeli_2018,lothringer,Kreidberg_w103}. Between 1 and 1.4 $\mu m$, both water and TiO/VO opacities are low, leading to a region where H-, TiO, and H$_2$O opacities have similar strength. Particularly, H- opacity fills the gap between the two water bands at 1.1 and 1.4 $\mu m$, effectively lowering the contrast between the top and the bottom of the bands. 

From Figure \ref{fig_eq_chem} we also note that the quick depletion of molecules in the atmosphere may introduce inaccuracies in the retrieval, as it assumes a single chemical abundance for the whole atmospheric pressure range. However, the chemical abundances of most molecules remain roughly constant for the pressure range that can be probed with our observations. Here, our isochemical retrievals suggest a thermal inversion which would be attributed to the absorption of TiO and VO at high altitudes. We could therefore expect the abundance of TiO and VO to differ significantly with pressure. However, the data quality is unlikely to support a retrieval with such complexity due to the narrow wavelength coverage but such complexities will need to be accounted for in the analysis of data from the next generation of facilities \citep{changeat}. \vspace{3mm}

\subsection{Previous Claims of Optical Absorbers}

Optical absorbers have been proposed as one of the leading theories as to why ultra hot Jupiters exhibit thermal inversions \citep[e.g.][]{Fortney_2008}. Hence, many atmospheric studies of these planets have been undertaken through both transmission and emission spectroscopy, with some planets studied through both methods.

WASP-19 b has been studied via transmission spectroscopy at near-infrared wavelengths with claims confirming and refuting the presence of TiO. The retrievals of the STIS G430L, G750L, WFC3 G141 and Spitzer IRAC observations suggest the presence of water at log(H$_2$O) $\approx$ -4 but show no evidence of optical absorbers \citep{sing,barstow_10planets,pinhas}. However, ground-based transits acquired with the European Southern Observatory’s Very Large Telescope (VLT), using the low-resolution FORS2 spectrograph (R$\sim$3000) which covers the entire visible-wavelength domain (0.43–1.04 $\mu m$), suggested the presence of TiO, to a confidence level of 7.7$\sigma$ \citep{sedahati_tio_wasp19}. However, \citet{espinoza_w19} found a featureless spectrum and argue the results of \citet{sedahati_tio_wasp19} are likely to be contaminated by stellar activity.

Evidence for a thermal inversion and optical absorbers has been seen of HAT-P-7\,b, which was first studied in emission during the commissioning program of Kepler when the satellite detected the eclipse as part of an optical phase curve \citep{borucki_hatp7}. This optical eclipse measurement was combined with Spitzer photometry over 3.5-8 $\mu m$ to infer the presence of a thermal inversion \citep{Christiansen_hatp7}, which was suggested due to the high flux ratio in the 4.5 $\mu m$ channel of Spitzer compared to the 3.6 $\mu m$ channel. Their chemical equilibrium models associated these emission features with CO, H$_2$O and CH$_4$. A thermal inversion was also reported to provide the best fit to these data by the atmospheric models of \citet{spiegel_hatp7} and \citet{madhu_hatp7}. All three studies noted that models without a thermal inversion could also explain the data, though only with a very high abundance of CH$_4$. More recently, \cite{mansfield_hatp7} obtained two eclipses using the HST WFC3 G141 grism. When combined with previous observations, it was found to be best fit with a thermal inversion due to optical absorbers, but at a low statistical significance when compared to a simpler blackbody fit.

Finally, some planets have been observed with both transit and eclipse spectroscopy. For instance, studies of WASP-33\,b have suggested the presence of Aluminium Oxide (AlO) in its transmission spectrum  \citep{von_essen_w33} while the WFC3 emission is best-fit by TiO and a thermal inversion \citep{Haynes_Wasp33b_spectrum_em}. Other studies using WFC3 G141 that have concluded optical absorbers may be present include WASP-121 b by \citet{Evans_wasp121_e1}. WASP-121 b has an equilibrium temperature of 2500 K and H$_2$O was detected at a 5$\sigma$ confidence with indications of absorption at high altitudes implying the presence of VO or TiO. The best-fit VO abundance was log(VO) = -3.5$^{+0.4}_{-0.6}$. Subsequently, observations of WASP-121 b with the G102 grism were taken and combined with the original data. In this further study, H$^-$ was included as an opacity source and the results were not consistent with the previously recovered VO abundance \citep{Evans_wasp121_e2}. \citet{bourrier} also performed a retrieval on the combined data, with the addition of data from TESS, and concluded VO abundance of log(VO) = -6.03$^{+0.50}_{-0.69}$, far lower than the initial retrieval on WFC3 G141 data. Additionally, the optical phase curve presented in \cite{daylan_tess_wasp121} suggested inefficient heat transport. This agrees with the work of \cite{Fortney_2008}, which postulated that the presence of optical absorbers would lead to, and require, large day-night temperature contrasts. However, \citet{merritt_w121} used high-resolution ground-based observations to place limits on the maximum abundances of TiO and VO in the terminator of WASP-121\,b to log(TiO) $<$ -9.26 and log(VO) $<$ -7.88. Nevertheless the authors of this study note that these upper bounds are degenerate with the cloud deck and scattering properties while also being limited by the accuracy of the VO line lists. Thus, the presence of these optical absorbers cannot be definitively ruled out as yet.

Therefore, our analysis here makes WASP-76\,b only the second ultra-hot Jupiter to be studied through both transmission and emission spectroscopy using WFC3 in scanning mode. In the analysis of WASP-121\,b's transmission and emission spectra by \citet{Evans_wasp121_t2, Evans_wasp121_e2}, chemical equilibrium models were used to fit the data. These suggested super-solar metallicities of 10-30 and 5-50x solar to 1$\sigma$ in transmission and emission respectively. These metallicity ranges provide H$_2$O abundances which are similar to those recovered here (10$^{-3}$-10$^{-4}$). The models from \cite{Parmentier_2018_w121photodiss} suggest that, for WASP-121\,b, $\sim 70\%$ of the H$_2$O in the 1.4$\mu m$ photosphere should be dissociated, compared to $\sim 50\%$ in WASP-76\,b. Additionally, the metallicity (Fe/H) of WASP-76 is greater than that of WASP-121 and both are above solar, at 0.366 and 0.12 respectively \citep{ehrenreich_wasp76, Evans_wasp121_e2}. We could therefore expect WASP-76\,b to have slightly more H$_2$O than WASP-121\,b but, while the best-fit solution agrees with this prediction, the 1$\sigma$ errors on the abundance are too large to be conclusive.

\subsection{Further Characterisation of WASP-76\,b}

\begin{figure}
    \centering
    \includegraphics[width = \columnwidth]{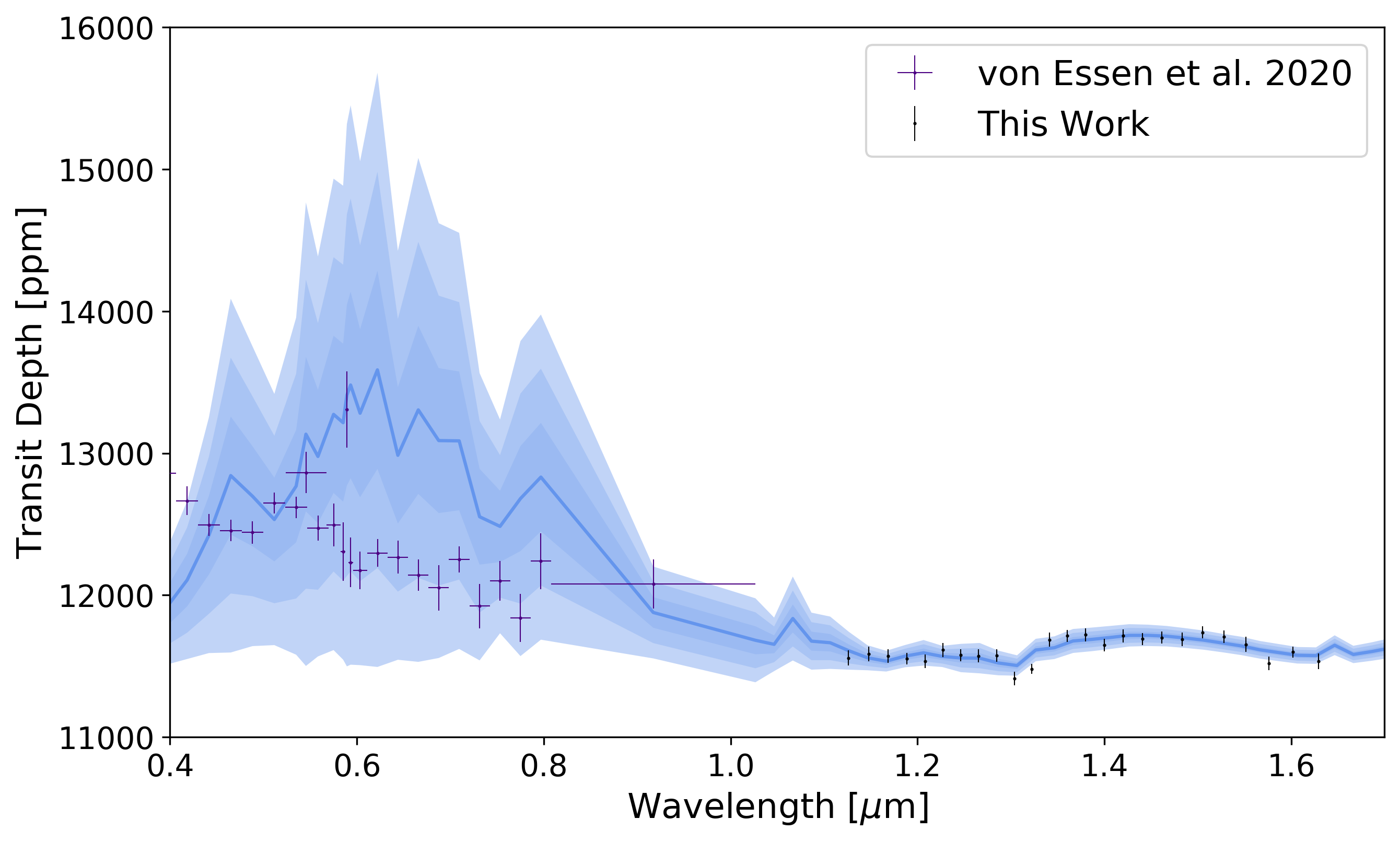}
    \caption{Comparison of the best-fit model to the WFC3 data and the STIS data from \citet{von_essen_w76}. Until a transmission spectrum is obtained with HST WFC3 G102 (0.8-1.1 $\mu m$), which would provide continuous wavelength coverage between STIS and WFC3 G141, the compatibility of the data sets cannot be ascertained.}
    \label{fig:stis_wfc3}
\end{figure}{}

The atmosphere of WASP-76\,b has been characterised in a number of other works. Notably, \citet{von_essen_w76} used HST STIS to study the transmission spectrum of WASP-76\,b. Hence, we extrapolated our best-fit model to the WFC3 data into the visible and over-plotted the data from \cite{von_essen_w76}. Figure \ref{fig:stis_wfc3} shows that, in the spectral region covered STIS, our uncertainties are very large. This is due to the wide range of abundances that TiO, VO and FeH could take, based on our analysis of WFC3 alone, and thus it is tempting to combine the data sets to reduce said uncertainty. However, without overlapping wavelength coverage, this is a dangerous pursuit at the best of times as the spectra could be offset due to the imperfect correction of instrument systematics, differing orbital parameters used in the fitting of the light curves, or temporal variations of the star-planet system. In this study, we have the additional complexity of a the third source contamination and differing methods in the removal of this stellar companion. For WASP-12\,b, which also suffered this issue, \cite{Kreidberg_wasp12} found the WFC3 data to be incompatible with that from STIS. Therefore we must, for now, resist the temptation to amalgamate data sets from multiple instruments. However, the addition of a transit observation with the G102 grism would provide continuous wavelength coverage from 0.3-1.7 $\mu m$, confirming the compatibility of the data and allowing the planet's terminator to be studied in far greater detail.

The acquisition of a secondary eclipse observation of WASP-76 b with the G102 grism of WFC3, which would extend the wavelength coverage into the red optical where emission bands due to species such as TiO, VO and FeH are more easily detectable, would further our knowledge of this planet and be valuable in providing additional evidence for, or refuting the presence of, TiO and in searching for other optical absorbers. 

\begin{figure*}
    \centering
    \includegraphics[width = 0.475\textwidth]{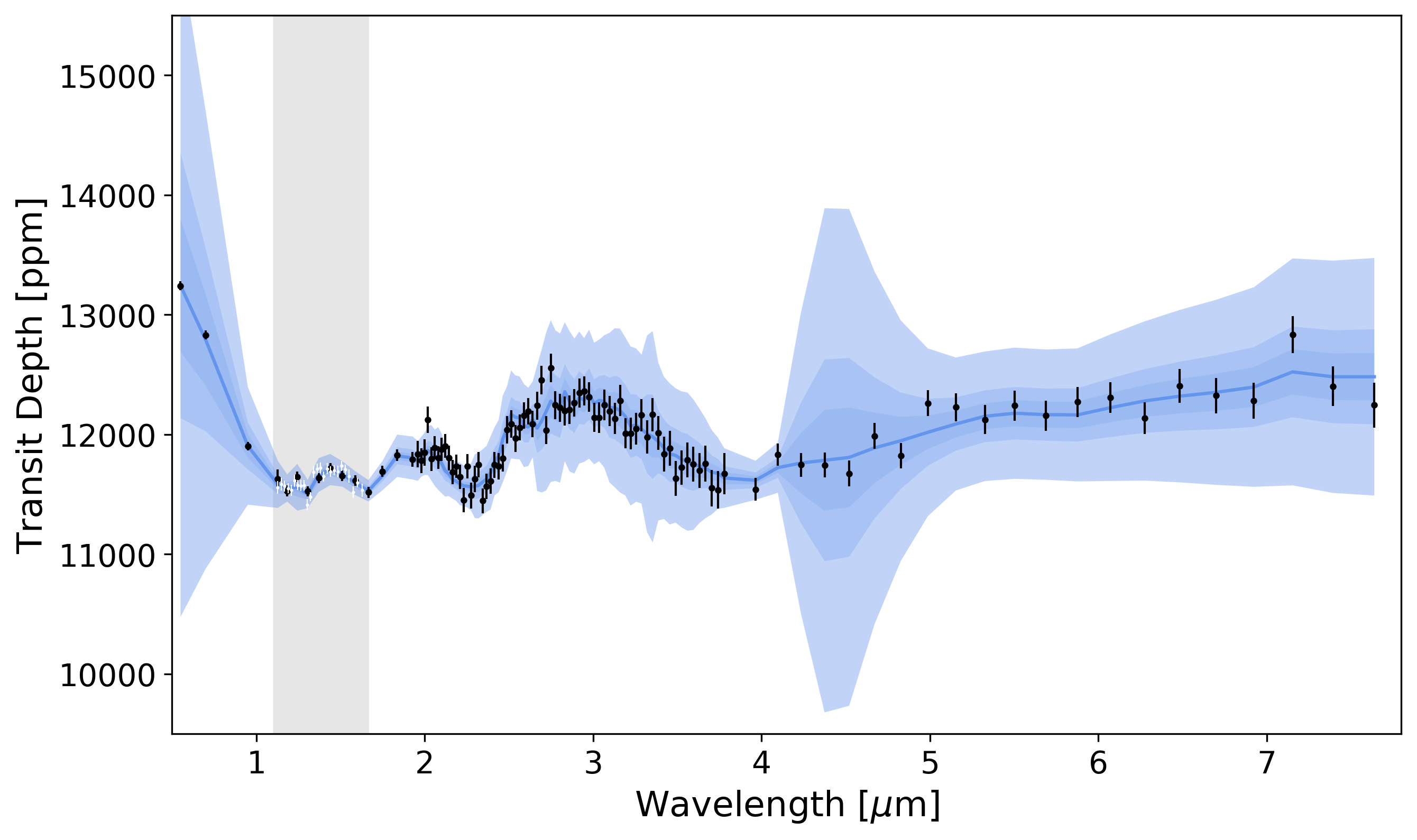}
     \includegraphics[width = 0.475\textwidth]{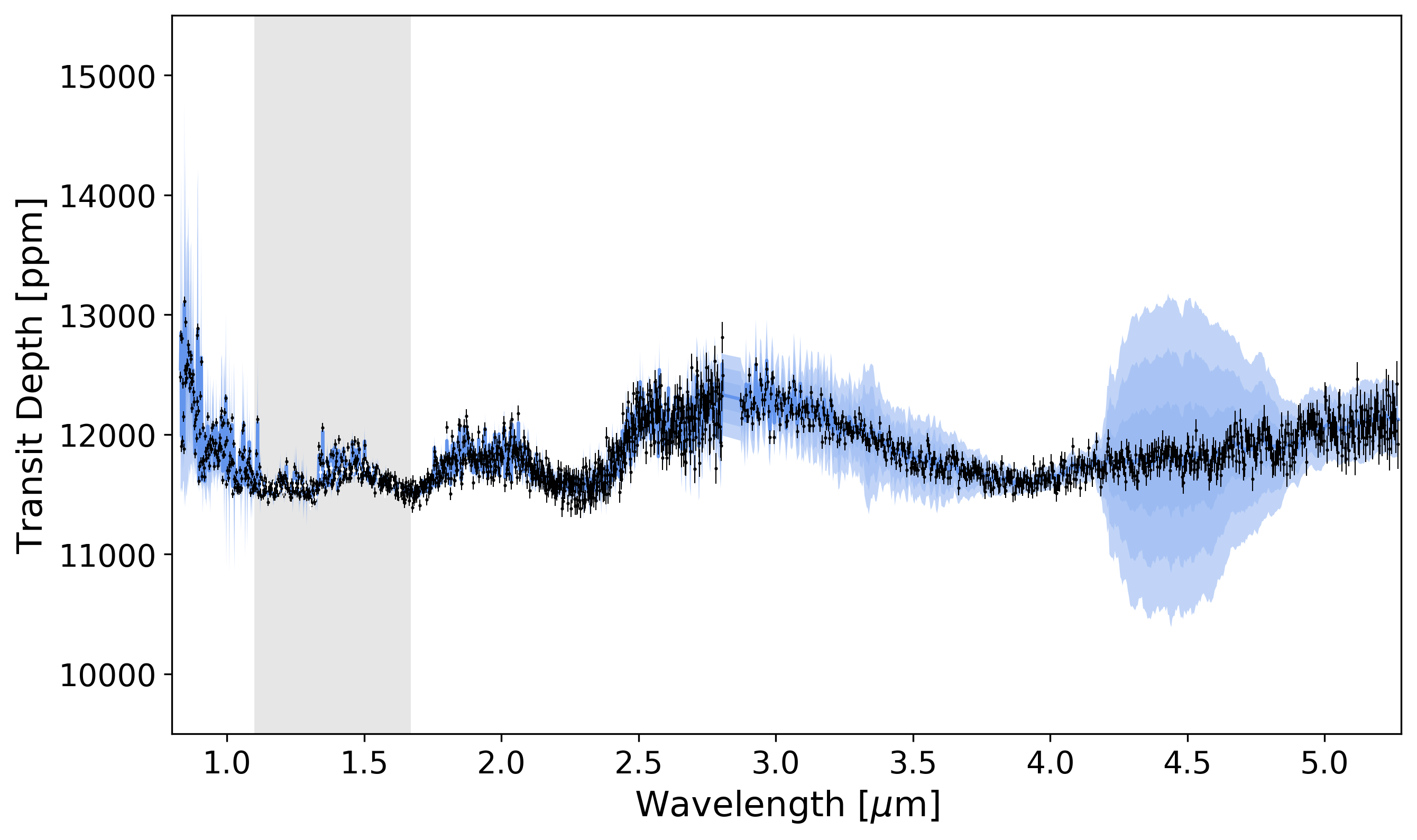}
    \includegraphics[width = 0.475\textwidth]{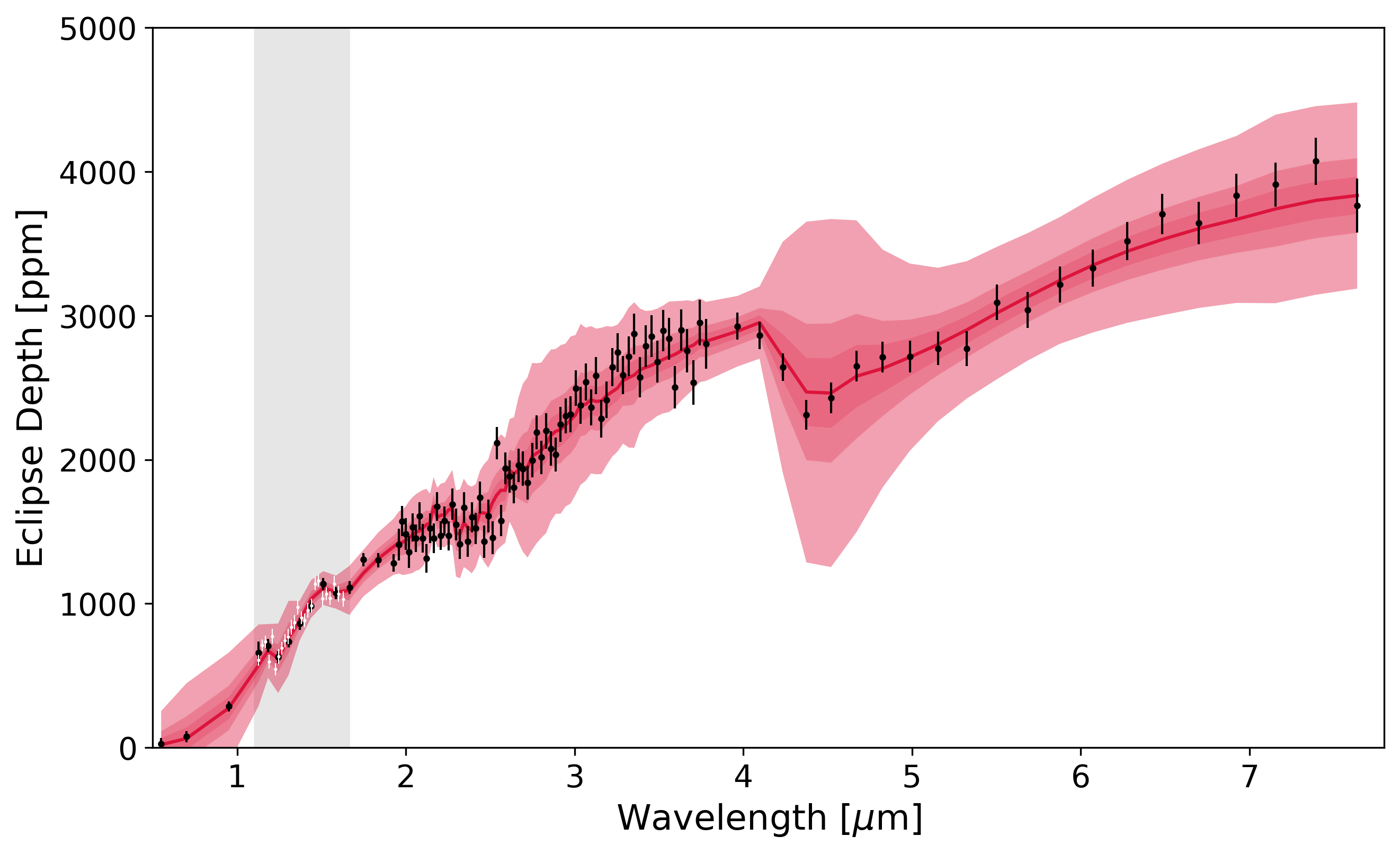}
    \includegraphics[width = 0.475\textwidth]{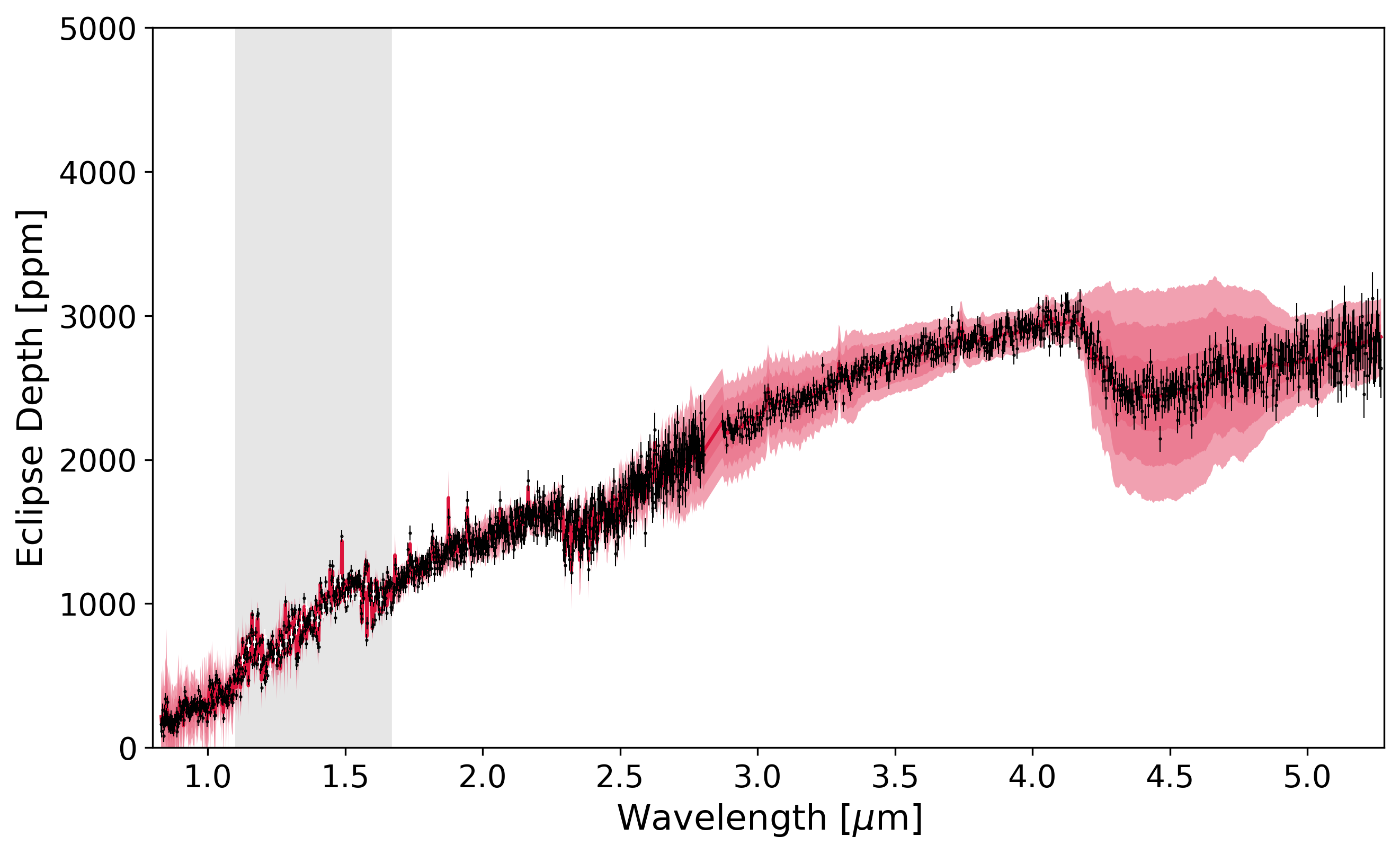}
    \caption{Simulated Ariel and JWST observations of the best-fit spectra from Hubble WFC3 observations. The Ariel spectra (left) are for a single observation at the native resolution of the instrumentation (i.e. Tier 3) while the JWST spectra (right) are for two observations, one using NIRISS G700XD and the second with NIRSpec G395M. The grey box indicates the wavelength range covered by the G141 grism and the data points from this study are shown in white. Note that, for all simulated observations, we have added Gaussian scatter.}
    \label{future_spec}
\end{figure*}{}

Future space telescopes JWST \citep{greene_jwst_exo}, Twinkle \citep{edwards_exo} and Ariel \citep{tinetti_ariel} will provide a far wider wavelength range. These missions will definitively move the exoplanet field from an era of detection into one of characterisation, allowing for the identification of the molecular species present and their chemical profile, insights into the atmospheric temperature profile and the detection and characterisation of clouds. Ariel, the ESA M4 mission due for launch in 2028, will conduct a survey of $\sim$1000 planets to answer the question: how chemically diverse are the atmospheres of exoplanets? WASP-76 b is an excellent target for study with Ariel \citep{edwards_ariel}, through both transmission and emission spectroscopy, and simulated error bars from \citet{mugnai} have been added to the best-fit spectra to showcase this in Figure \ref{future_spec}. Additionally, ExoWebb \citep{exowebb} has been used to simulate the capability of JWST for studying this planet. For both facilities, the predicted error bars are far smaller than the 1$\sigma$ uncertainty in the best-fit spectrum to the Hubble WFC3 data and thus they will allow for far tighter constraints on molecular constituents to be imposed. 

\section{Conclusions}

Both the transmission and emission spectra of WASP-76 b, obtained by Hubble WFC3, have been analysed. We have used open-source codes to reduce the data, remove contamination from a close stellar companion, and fit the final spectra. The transmission spectrum exhibits a large water feature while the dayside of the planet shows strong evidence for titanium oxide, as well as water, and is best-fit by an atmospheric thermal inversion. The abundances retrieved closely match those from chemical equilibrium models. However, further observations with Hubble, or future space-based facilities, will result in a better understanding of the chemical constituents of the atmosphere and help refine the models presented here.

\vspace{5mm}
\textbf{Acknowledgements:} This work was realised as part of ARES, the Ariel Retrieval Exoplanet School, in Biarritz in September 2019. The school was organised by Jean-Philippe Beaulieu, Angelos Tsiaras and Ingo Waldmann with the financial support of CNES. The publicly available observations presented here were taken as part of proposals 14260 and 14767, led by Drake Deming and David Sing respectively. These were obtained from the Hubble Archive which is part of the Mikulski Archive for Space Telescopes.

BE, QC, MM, AT and IW are funded through the ERC Consolidator grant ExoAI (GA 758892) and the STFC grants ST/P000282/1, ST/P002153/1, ST/S002634/1 and ST/T001836/1. RB is a Ph.D. fellow of the Research Foundation--Flanders (FWO). NS acknowledges the support of the IRIS-OCAV, PSL. MP acknowledges support by the European Research Council under Grant Agreement ATMO 757858 and by the CNES. WP and TZ have received funding from the European Research Council (ERC) under the European Union's Horizon 2020 research and innovation programme (grant agreement no. 679030/WHIPLASH). OV thanks the CNRS/INSU Programme National de Plan\'etologie (PNP) and CNES for funding support. JPB acknowledges the support of the University of Tasmania through the UTAS Foundation and the endowed Warren Chair in Astronomy, Rodolphe Cledassou, Pascale Danto and Michel Viso (CNES).

We are grateful to John Southworth for informing us on the presence, and the characteristics, of the stellar companion. We also thank Paul Molli\`ere, for helpful comments and for providing the latest updates to petitCODE, and Karan Molaverdikhani for providing atomic and ion opacities for our petitCODE simulations. 

Finally, we thank our anonymous referee for prompt, insightful comments which led to the improvement of the manuscript.

\vspace{3mm}
\textbf{Software:} Iraclis \citep{tsiaras_hd209}, TauREx3 \citep{al-refaie_taurex3}, Wayne \citep{varley}, petitCODE \citep{molliere_petitcode}, pylightcurve \citep{tsiaras_plc}, ExoTETHyS \citep{morello_exotethys}, ArielRad \citep{mugnai}, ExoWebb \citep{exowebb}, Astropy \citep{astropy}, h5py \citep{hdf5_collette}, emcee \citep{emcee}, Matplotlib \citep{Hunter_matplotlib}, Multinest \citep{Feroz_multinest}, Pandas \citep{mckinney_pandas}, Numpy \citep{oliphant_numpy}, SciPy \citep{scipy}.

\begin{table*}
    \centering
    \begin{tabular}{ccccccccccc}\hline \hline
Wavelength & Bandwidth & Correction & \multicolumn{4}{c}{Transit} & \multicolumn{4}{c}{Eclipse}\\
$[ \mu m]$ & $[ \mu m]$ & Factor & Depth [\%] & ${\chi}^2$ & $\bar{\sigma}$ & AC & Depth [\%] & ${\chi}^2$ & $\bar{\sigma}$ & AC \\\hline
1.12625 & 0.0219 & 1.080007 & 1.1557 $\pm$ 0.0054 & 1.07 & 1.29 & 0.27 & 0.0607 $\pm$ 0.0039 & 1.06 & 0.90 & 0.18 \\
1.14775 & 0.0211 & 1.081612 & 1.1585 $\pm$ 0.0052 & 1.07 & 1.24 & 0.05 & 0.0711 $\pm$ 0.0040 & 1.06 & 0.93 & 0.07 \\
1.16860 & 0.0206 & 1.083408 & 1.1570 $\pm$ 0.0047 & 1.07 & 1.14 & 0.20 & 0.0736 $\pm$ 0.0043 & 1.06 & 0.99 & 0.23 \\
1.18880 & 0.0198 & 1.084441 & 1.1551 $\pm$ 0.0041 & 1.07 & 1.00 & 0.17 & 0.0597 $\pm$ 0.0048 & 1.06 & 1.14 & 0.21 \\
1.20835 & 0.0193 & 1.085204 & 1.1534 $\pm$ 0.0048 & 1.07 & 1.11 & 0.26 & 0.0772 $\pm$ 0.0053 & 1.06 & 1.24 & 0.09 \\
1.22750 & 0.0190 & 1.086487 & 1.1614 $\pm$ 0.0050 & 1.07 & 1.20 & 0.11 & 0.0544 $\pm$ 0.0041 & 1.08 & 1.08 & 0.18 \\
1.24645 & 0.0189 & 1.087721 & 1.1578 $\pm$ 0.0042 & 1.07 & 1.03 & 0.29 & 0.0633 $\pm$ 0.0050 & 1.06 & 1.19 & 0.08 \\
1.26550 & 0.0192 & 1.089421 & 1.1572 $\pm$ 0.0047 & 1.07 & 1.14 & 0.08 & 0.0692 $\pm$ 0.0042 & 1.06 & 0.98 & 0.15 \\
1.28475 & 0.0193 & 1.091716 & 1.1574 $\pm$ 0.0044 & 1.07 & 1.07 & 0.32 & 0.0742 $\pm$ 0.0047 & 1.06 & 1.19 & 0.06 \\
1.30380 & 0.0188 & 1.091428 & 1.1414 $\pm$ 0.0048 & 1.07 & 1.17 & 0.17 & 0.0770 $\pm$ 0.0053 & 1.06 & 1.27 & 0.18 \\
1.32260 & 0.0188 & 1.092315 & 1.1480 $\pm$ 0.0036 & 1.07 & 0.90 & 0.24 & 0.0836 $\pm$ 0.0052 & 1.06 & 1.20 & 0.17 \\
1.34145 & 0.0189 & 1.093736 & 1.1686 $\pm$ 0.0050 & 1.07 & 1.20 & 0.09 & 0.0870 $\pm$ 0.0050 & 1.06 & 1.17 & 0.18 \\
1.36050 & 0.0192 & 1.095211 & 1.1713 $\pm$ 0.0043 & 1.07 & 1.02 & 0.21 & 0.0976 $\pm$ 0.0052 & 1.06 & 1.19 & 0.13 \\
1.38005 & 0.0199 & 1.096720 & 1.1721 $\pm$ 0.0046 & 1.07 & 1.06 & 0.08 & 0.0903 $\pm$ 0.0044 & 1.06 & 1.01 & 0.16 \\
1.40000 & 0.0200 & 1.097740 & 1.1649 $\pm$ 0.0044 & 1.07 & 1.05 & 0.10 & 0.0882 $\pm$ 0.0048 & 1.06 & 1.09 & 0.28 \\
1.42015 & 0.0203 & 1.097564 & 1.1714 $\pm$ 0.0046 & 1.07 & 1.09 & 0.25 & 0.0955 $\pm$ 0.0051 & 1.06 & 1.13 & 0.23 \\
1.44060 & 0.0206 & 1.099283 & 1.1691 $\pm$ 0.0042 & 1.07 & 1.01 & 0.15 & 0.0988 $\pm$ 0.0048 & 1.06 & 1.06 & 0.05 \\
1.46150 & 0.0212 & 1.100529 & 1.1701 $\pm$ 0.0043 & 1.07 & 0.97 & 0.12 & 0.1139 $\pm$ 0.0045 & 1.06 & 0.99 & 0.08 \\
1.48310 & 0.0220 & 1.102016 & 1.1689 $\pm$ 0.0048 & 1.07 & 1.08 & 0.08 & 0.1160 $\pm$ 0.0044 & 1.06 & 0.95 & 0.09 \\
1.50530 & 0.0224 & 1.103614 & 1.1737 $\pm$ 0.0042 & 1.07 & 1.00 & 0.04 & 0.1035 $\pm$ 0.0051 & 1.06 & 1.14 & 0.01 \\
1.52800 & 0.0230 & 1.107372 & 1.1707 $\pm$ 0.0042 & 1.07 & 1.00 & 0.09 & 0.1067 $\pm$ 0.0045 & 1.06 & 1.00 & 0.02 \\
1.55155 & 0.0241 & 1.109843 & 1.1652 $\pm$ 0.0054 & 1.07 & 1.23 & 0.06 & 0.1038 $\pm$ 0.0048 & 1.06 & 1.06 & 0.07 \\
1.57625 & 0.0253 & 1.110741 & 1.1518 $\pm$ 0.0048 & 1.07 & 1.10 & 0.18 & 0.1138 $\pm$ 0.0053 & 1.06 & 1.15 & 0.13 \\
1.60210 & 0.0264 & 1.113385 & 1.1598 $\pm$ 0.0040 & 1.07 & 0.90 & 0.02 & 0.1067 $\pm$ 0.0056 & 1.06 & 1.23 & 0.23 \\
1.62945 & 0.0283 & 1.114973 & 1.1535 $\pm$ 0.0056 & 1.07 & 1.17 & 0.14 & 0.1031 $\pm$ 0.0054 & 1.06 & 1.15 & 0.13 \\ \hline \hline
    \end{tabular}
    \caption{Corrected transmission and emission spectra derived here along with the Chi-squared ($\chi^2$), the standard deviation of the residuals with respect to the photon noise ($\bar{\sigma}$) and the auto-correlation (AC) for the spectral light curve fits. Note that the correction factor has already been applied to the transit and eclipse depths.}
    \label{tab:spectra}
\end{table*}

\clearpage
\bibliographystyle{aasjournal}
\bibliography{main}

\end{document}